\documentclass[11pt]{article}
\usepackage{epsfig}
\def\nn{\nonumber}
\def\be{\begin{equation}}
\def\ee{\end{equation}}
\newcommand{\bea}{\begin{eqnarray}}
\newcommand{\eea}{\end{eqnarray}}
\newcommand{\bdm}{\begin{displaymath}}
\newcommand{\edm}{\end{displaymath}}
\long\def\symbolfootnote[#1]#2{\begingroup%
\def\thefootnote{\fnsymbol{footnote}}\footnote[#1]{#2}\endgroup}

\def\sq2{\sqrt{2}}
\def\drbar{\overline{\rm DR}}
\def\msbar{\overline{\rm MS}}

\def\smallsm{\scriptscriptstyle{\rm SM}}
\def\tb{\tan\beta}
\def\cbe{c_\beta}
\def\sbe{s_\beta}
\def\gl{\tilde{g}}
\def\mg{m_{\gl}}
\def\g{\mg^2}
\def\gq{\mg^4}

\def\hsm{H_{\scriptscriptstyle{\rm SM}}}

\def\x1g{x_{1}}

\newcommand{\as}{\alpha_s}
\newcommand{\oas}{{\cal O}(\as)}
\newcommand{\smallz}{{\scriptscriptstyle Z}} %  a smaller Z
\newcommand{\smallw}{{\scriptscriptstyle W}} %
\newcommand{\smallr}{{\scriptscriptstyle R}} %
\newcommand{\smalla}{{\scriptscriptstyle A}} %
\newcommand{\mz}{m_\smallz}
\newcommand{\mw}{m_\smallw}

\newcommand{\ma}{m_\smalla}
\newcommand{\muF}{\mu_{\scriptscriptstyle F}}
\newcommand{\muR}{\mu_\smallr}
\newcommand{\HA}{{\mathcal H}_\smalla}
\newcommand{\HAt}{{\mathcal K}_{t \tilde t \gl}}
\newcommand{\HAb}{{\mathcal K}_{b \tilde b \gl}}
\def\dmbsusy{\frac{(\delta m_b)}{\mb}^{\scriptscriptstyle SUSY}}

\def\mt{m_t}
\def\stu{\tilde{t}_1}
\def\std{\tilde{t}_2}

\def\t{\mt^2}

\def\tu{m_{\tilde{t}_1}^2}
\def\td{m_{\tilde{t}_2}^2}
\def\ti{m_{\tilde{t}_i}^2}

\def\tiq{m_{\tilde{t}_i}^4}
\def\sdt{s_{2\theta_t}}

\def\mb{m_b}
\def\sbu{\tilde{b}_1}
\def\sbd{\tilde{b}_2}

\def\bu{m_{\tilde{b}_1}^2}
\def\bd{m_{\tilde{b}_2}^2}
\def\bi{m_{\tilde{b}_i}^2}

\def\sdb{s_{2\theta_b}}

\newenvironment{appendletterA}
 {
  \setcounter{section}{0}
  \setcounter{equation}{0}
  
 }{
 }

\begin{document}

\begin{titlepage}

%\today

{\flushright{
        \begin{minipage}{5cm}
          RM3-TH/11-05
        \end{minipage}        }

}
\renewcommand{\thefootnote}{\fnsymbol{footnote}}
\vskip 2cm
\begin{center}
\boldmath
{\LARGE\bf NLO QCD corrections to pseudoscalar \\[7pt]
Higgs production in the MSSM}\unboldmath
\vskip 1.cm
{\Large{G.~Degrassi$^{a}$, S.~Di~Vita$^{a}$ and P.~Slavich$^{b}$}}
\vspace*{8mm} \\
{\sl ${}^a$
    Dipartimento di Fisica, Universit\`a di Roma Tre and  INFN, Sezione di
    Roma Tre \\
    Via della Vasca Navale~84, I-00146 Rome, Italy}
\vspace*{2.5mm}\\
{\sl ${}^b$  LPTHE, 4, Place Jussieu, F-75252 Paris,  France}
\end{center}
\symbolfootnote[0]{{\tt e-mail:}}
\symbolfootnote[0]{{\tt degrassi@fis.uniroma3.it}}
\symbolfootnote[0]{{\tt divita@fis.uniroma3.it}}
\symbolfootnote[0]{{\tt slavich@lpthe.jussieu.fr}}

\vskip 0.7cm

\begin{abstract}
  We present a calculation of the two-loop quark-squark-gluino
  contributions to pseudoscalar Higgs boson production via gluon
  fusion in the MSSM. We regularize the loop integrals using the
  Pauli-Villars method, and obtain explicit and compact analytic
  results based on an expansion in the heavy particle masses. Our
  results -- valid when the pseudoscalar Higgs boson is lighter than
  squarks and gluinos -- can be easily implemented in computer codes
  for an efficient and accurate determination of the
  pseudoscalar production cross section.

\end{abstract}
\vfill
\end{titlepage}    
\setcounter{footnote}{0}

%%%%%%%%%%%%%%%%%%%%%%%%%%%%%%%%%%%%%%%%%%%%%%%%%%%%%%%%%%%%%%%%%%%%%%%%%%%%%%%

\section{Introduction}

With the coming into operation of the Large Hadron Collider (LHC), a
new era has begun in the search for the Higgs boson(s). At the LHC the
main production mechanism for the Standard Model (SM) Higgs boson,
$\hsm$, is the loop-induced gluon fusion mechanism \cite{H2gQCD0}, $gg
\to \hsm$, where the coupling of the gluons to the Higgs is mediated
by loops of colored fermions, primarily the top quark. The knowledge
of this process in the SM includes the full next-to-leading order
(NLO) QCD corrections \cite{H2gQCD1,SDGZ,HK0}, the
next-to-next-to-leading order (NNLO) QCD corrections \cite{H2gQCD2}
including finite top mass effects \cite{H2gQCD3}, soft-gluon
resummation effects \cite{bd4}, an estimate of the
next-to-next-to-next-to-leading order (NNNLO) QCD effects
\cite{Moch:2005ky} and also the first-order electroweak corrections
\cite{DjG,ABDV0,APSU}.

The Higgs sector of the Minimal Supersymmetric extension of the
Standard Model (MSSM) consists of two $SU(2)$ doublets, $H_1$ and
$H_2$, whose relative contribution to electroweak symmetry breaking is
determined by the ratio of vacuum expectation values of their neutral
components, $\tb\equiv v_1/v_2$. The spectrum of physical Higgs bosons
is richer than in the SM, consisting of two neutral CP-even bosons,
$h$ and $H$, one neutral CP-odd boson, $A$, and two charged scalars,
$H^\pm$. The couplings of the MSSM Higgs bosons to matter fermions
differ from those of the SM Higgs, and they can be considerably enhanced
(or suppressed) depending on $\tb$. As in the SM case, the
gluon-fusion process is one of the most important production
mechanisms for the neutral Higgs bosons, whose couplings to the gluons
are mediated by top and bottom quarks and their supersymmetric
partners, the stop and sbottom squarks.

In the case of the CP-even bosons $h$ and $H$ the gluon-fusion cross
section in the MSSM is known at the NLO in QCD.\footnote{First results
  for the NNLO contributions in the limit of degenerate superparticle
  masses were presented in ref.~\cite{NNLOSUSY}.}  The contributions
arising from diagrams with quarks and gluons can be obtained from the
corresponding SM results with an appropriate rescaling of the
Higgs-quark couplings. The contributions arising from diagrams with
squarks and gluons were first computed under the approximation of
vanishing Higgs mass in ref.~\cite{Dawson:1996xz}.  The complete
top/stop contributions, including the effects of stop mixing and of
the two-loop diagrams involving gluinos, were computed under the same
approximation in ref.~\cite{HS}, and the result was cast in a compact
analytic form in ref.~\cite{DS}. Later calculations aimed at the
inclusion of the full Higgs-mass dependence in the squark-gluon
contributions, which are now known in a closed analytic form
\cite{babis1,ABDV,MS,BDV}.

The approximation of vanishing Higgs mass in the contributions of
two-loop diagrams allows for compact analytic results that can be
implemented in computer codes for a fast and efficient evaluation of
the Higgs production cross section. For what concerns the top-gluon
contributions, the effect of such approximation on the result for the
cross section has been shown \cite{KLS,BDV} to be limited to a few
percent, as long as the Higgs mass is below the threshold for creation
of the massive particles running in the diagrams (in this case, the
top quarks). While this condition may also apply to the two-loop
diagrams involving top, stop and gluino, it obviously does not apply
to the corresponding diagrams involving the bottom quark, whose
contribution can be relevant for large values of $\tb$. For the latter
diagrams the dependence on the Higgs mass should in principle be
retained, which has proved a rather daunting task. A calculation of
the full quark-squark-gluino contributions via a combination of
analytic and numerical methods was presented in ref.~\cite{babis2}
(see also ref.~\cite{spiraDb}), but neither explicit analytic results
nor a public computer code have been made available so far. However,
ref.~\cite{noibot} presented an evaluation of the
bottom-sbottom-gluino diagrams based on an asymptotic expansion in the
large supersymmetric masses that is valid up to and including terms of
${\cal O}(\mb^2/m_\phi^2)$, ${\cal O}(\mb/M)$ and ${\cal
  O}(\mz^2/M^2)$, where $m_\phi$ denotes a Higgs boson mass and $M$
denotes a generic superparticle mass. This expansion should provide a
good approximation to the full result, at least comparable to the one
obtained for the top-stop-gluino diagrams, as long as the Higgs boson
mass is below all the heavy-particle thresholds. An independent
calculation of the bottom-sbottom-gluino contributions, restricted to
the limit of a degenerate superparticle mass spectrum, was also
presented in ref.~\cite{hhm}, confirming the results of
ref.~\cite{noibot}.

In the case of the CP-odd boson $A$ the calculation of the production
cross section is somewhat less advanced. Due to the structure of the
$A$-boson coupling to squarks, only loops of top and bottom quarks
contribute to the cross section at LO, with the bottom loops being
dominant for even moderately large values of $\tb$. In the limit of
vanishing $A$-boson mass, $\ma$, the contributions from diagrams with
quarks and gluons were computed at NLO in ref.~\cite{AQCD} and at NNLO
in ref.~\cite{CKSB} (see also ref.~\cite{CM}). For arbitrary values of
$\ma$ the NLO contributions arising from two-loop diagrams with quarks
and gluons, as well as from one-loop diagrams with emission of a real
parton, were computed in ref.~\cite{SDGZ}. Supersymmetric particles
contribute to the cross section at NLO through two-loop diagrams
involving quarks, squarks and gluinos. The top-stop-gluino
contributions were computed in ref.~\cite{HH} in the limit of
vanishing $\ma$. The analytic result for generic values of the stop
and gluino masses was deemed too voluminous to be explicitly displayed
in ref.~\cite{HH}, and was instead made available in the fortran code
{\tt evalcsusy.f} \cite{HS}. On the other hand, the two-loop
bottom-sbottom-gluino contributions, which can be relevant for large
values of $\tb$, have never been directly computed so far.

In this paper we aim to reduce the gap in accuracy between the
available NLO calculations of the production cross sections for CP-odd
and CP-even Higgs bosons of the MSSM, exploiting the techniques we
developed for computing the top-stop-gluino \cite{DS} and
bottom-sbottom-gluino \cite{noibot} contributions in the CP-even case.
In particular, we present an evaluation of the two-loop
top-stop-gluino contributions to the pseudoscalar production cross
section valid up to and including terms of ${\cal O}(\ma^2/\mt^2)$ and
${\cal O}(\ma^2/M^2)$. We show how the terms of order zero in $\ma^2$
can be cast in an extremely compact analytic form, fully equivalent to
the result of ref.~\cite{HH}, and we investigate the effect of the
first-order terms. We also evaluate the same contributions via an
asymptotic expansion in the large superparticle masses, valid up to
and including terms of ${\cal O}(\ma^2/M^2)$ and ${\cal
  O}(\mt^2/M^2)$. While the latter result is valid for $\mt, \ma \ll
M$ but does not assume a hierarchy between $\mt$ and $\ma$, the former
is expected to provide a better approximation in the region with $\ma
< \mt$ and relatively light superparticles, $M \simeq \mt$.  As a
byproduct, we also obtain a result for the bottom-sbottom-gluino
contributions valid up to and including terms of ${\cal
  O}(\mb^2/\ma^2)$ and ${\cal O}(\mb/M)$. Finally, we compare our
results for the bottom-sbottom-gluino contributions to both CP-even
and CP-odd Higgs production cross sections with those obtained in the
effective-Lagrangian approximation of refs.~\cite{effL,GHS}.

A non-trivial technical issue that arises in the calculation of the
pseudoscalar production cross section is the treatment of the Dirac
matrix $\gamma_5$ -- an intrinsically four-dimensional object --
within regularization methods defined in a number of dimensions $n_d =
4-2\epsilon$. The original calculation of the two-loop quark-gluon
contributions of ref.~\cite{SDGZ} was performed in Dimensional
Regularization (DREG), employing the 't Hooft-Veltman (HV)
prescription \cite{HV} for the $\gamma_5$ matrix and introducing a
finite multiplicative renormalization factor \cite{larin} to restore
the Ward identities. In ref.~\cite{HH} the calculation of the
top-gluon and top-stop-gluino contributions to the Wilson coefficient
in the relevant effective Lagrangian was performed both in DREG and in
Dimensional Reduction (DRED), which, differently from DREG, preserves
supersymmetry (SUSY). The latter method does not require the
introduction of finite renormalization factors, but it involves
additional subtleties concerning the treatment of the Levi-Civita
symbol $\varepsilon_{\mu\nu\rho\sigma}$.

In our calculation of the quark-squark-gluino contributions we avoided
all problems related to the treatment of $\gamma_5$ by employing the
Pauli-Villars regularization (PVREG) method. Being defined in four
dimensions, the PVREG method respects both SUSY and the chiral
symmetry, therefore no symmetry-restoring renormalization factors need
to be introduced. We tested our implementation of PVREG by computing
the top-gluon contributions via an asymptotic expansion in the top
quark mass, and recovering the result obtained in DREG in
refs.~\cite{SDGZ,ABDV}. As a further cross check, we also computed the
quark-squark-gluino contributions using the DREG procedure outlined in
ref.~\cite{larin}, and found agreement with the result that we
obtained in PVREG.

The paper is organized as follows: in section \ref{sec:general} we
summarize general results on the cross section for pseudoscalar Higgs
boson production via gluon fusion. In section \ref{sec:calc} we
outline our implementation of the PVREG method.  Section
\ref{sec:2loopres} contains our explicit results for the NLO
contributions arising from both top-stop-gluino and
bottom-sbottom-gluino diagrams, as well as a discussion of suitable
renormalization schemes for the bottom contributions and a comparison
with the results obtained in the effective-Lagrangian
approximation. In section \ref{sec:num} we assess the validity of the
expansion in powers of $\ma^2$ in the top contributions, and discuss
the numerical relevance of the different NLO contributions. In the
last section we present our conclusions.  We also include, for
completeness, an Appendix in which we present the NLO contributions
from one-loop diagrams with emission of a real parton.

\vfill
%%%%%%%%%%%%%%%%%%%%%%%%%%%%%%%%%%%%%%%%%%%%%%%%%%%%%%%%%%%%%%%%%%%%%%%%%%%%
\newpage

\section{Pseudoscalar Higgs boson production via gluon fusion at NLO}
\label{sec:general}

In this section we recall for completeness some general results on
pseudoscalar Higgs boson production via gluon fusion.  The hadronic
cross section at center-of-mass energy $\sqrt{s}$ can be written as
\be
\sigma(h_1 + h_2 \to A+X)  \,=\, 
          \sum_{a,b}\int_0^1 dx_1 dx_2 \,\,f_{a,h_1}(x_1,\muF)\,
         f_{b,h_2}(x_2,\muF)  \times
\int_0^1 dz~ \delta \left(z-\frac{\tau_\smalla}{x_1 x_2} \right)
\hat\sigma_{ab}(z)~,
\label{sigmafull}
\ee
where $\tau_\smalla= \ma^2/s$, $\muF$ is the factorization scale,
$f_{a,h_i}(x,\muF)$ the parton density of the colliding hadron $h_i$
for the parton of type $a$ (for $a = g,q,\bar{q}$), and
$\hat\sigma_{ab}$ the cross section for the partonic subprocess $ ab
\to A +X$ at the center-of-mass energy $\hat{s}=x_1 \,x_2\,
s=\ma^2/z$. The partonic cross section can be written in terms of the LO
contribution $\sigma^{(0)}$ and a coefficient function $G_{ab}(z)$ as
\be
\hat\sigma_{ab}(z)=
\sigma^{(0)}\,z \, G_{ab}(z) \, .
\label{Geq}
\ee
The LO term can be written as
\be
\sigma^{(0)}  =  
\frac{G_\mu \,\alpha_s^2 (\muR)  }{128\, \sqrt{2} \, \pi}\,
\left|\HA^{1\ell} \right|^2~,
\label{ggh}
\ee
where $G_\mu$ is the muon decay constant and $\alpha_s(\muR)$ is the
strong gauge coupling expressed in the $\overline{\rm MS}$
renormalization scheme at the scale $\muR$. $\HA$ is the form factor
for the coupling of the pseudoscalar $A$ with two gluons, which we
decompose in one- and two-loop parts as
\be
\HA ~=~ \HA^{1\ell} ~+~ \frac{\alpha_s}{\pi} \, \HA^{2\ell}
~+~{\cal O}(\alpha_s^2)~.
\label{Hdec}
\ee

Due to the structure of the pseudoscalar coupling to squarks (see
section \ref{sec:2loopres}), only diagrams involving top or bottom
quarks contribute to the one-loop form factor $\HA^{1\ell}$. The
latter can be decomposed into top and bottom contributions as
\be
\HA^{1\ell} ~=~ T_F \,\left[\cot\beta\,{\cal K}^{1\ell}(\tau_t) 
+ \tan\beta\,{\cal K}^{1\ell}(\tau_b)\right]~,
\label{H1lp}
\ee
where $T_F=1/2$ is a color factor, $\tau_q = 4\,m_q^2/\ma^2\,$,  and
\be
{\cal K}^{1\ell}(\tau) = 
\frac{\tau}{2}\,\ln^2\left(\frac{\sqrt{1- \tau}-1}{\sqrt{1-\tau}+1}\right)~.
\label{K1l}
\ee
We recall the behavior of ${\cal K}^{1\ell}$ in the limit in which the
pseudoscalar mass is much smaller or much larger than twice the mass
of the particle running in the loop.  In the first case, i.e.~$\tau\gg
1$, which may apply to the top contribution if $\ma$ is relatively
small,
\be
{\cal K}^{1\ell}(\tau) 
~\longrightarrow~ -2 -\frac{2}{3\tau} ~+~ {\cal O}(\tau^{-2})~,
\label{K1lim1}
\ee
while in the opposite case, i.e.~$\tau\ll 1$, which is relevant for
the bottom contribution,
\be
{\cal K}^{1\ell}(\tau) 
~\longrightarrow~ \frac{\tau}{2}\,\ln^2 (\frac{-4}{\tau}) 
~+~ {\cal O}(\tau^2)~.
\label{K1lim2}
\ee
The analytic continuation of ${\cal K}^{1\ell}(\tau)$ corresponds to
the replacement $\ma^2 \rightarrow \ma^2 + i \epsilon$~, thus the
imaginary part of eq.~(\ref{K1lim2}) can be recovered via the
replacement $\ln(-4/\tau) \rightarrow \ln(4/\tau) - i\pi$.

The coefficient function $G_{ab}(z)$ in eq.~(\ref{Geq}) can be
decomposed, up to NLO terms, as
\be
G_{a b}(z)  ~=~  G_{a b}^{(0)}(z) 
~+~ \frac{\alpha_s}{\pi} \, G_{a b}^{(1)}(z) ~+~{\cal O}(\alpha_s^2)\, ,
\label{Gdec}
\ee
with the LO contribution given only by the gluon-fusion channel:
\bea
G_{a b}^{(0)}(z) & = & \delta(1-z) \,\delta_{ag}\, \delta_{bg} \, .
\eea
The NLO terms include, besides the $gg$ channel, also the one-loop
induced $gq$ and $q \bar{q}$ channels:
\bea
\label{ggg}
G_{g g}^{(1)}(z) & = & \delta(1-z) \left[C_A \, \frac{~\pi^2}3 
\,+ \,\beta_0 \, \ln \left( \frac{\muR^2}{\muF^2} \right) 
 \,+ \,2\,{\rm Re}\left(\frac{\HA^{2\ell}}{\HA^{1\ell}} \right)  \right]  \nn \\
&+ &  P_{gg} (z)\,\ln \left( \frac{\hat{s}}{\muF^2}\right) +
    C_A\, \frac4z \,(1-z+z^2)^2 \,{\cal D}_1(z) +  C_A\, {\cal R}_{gg}  \, , 
\eea

\be
G_{q \bar{q}}^{(1)}(z) ~=~   {\cal R}_{q \bar{q}} \, , ~~~~~~~~~~~
G_{q g}^{(1)}(z) ~=~  P_{gq}(z) \left[ \ln(1-z) + 
 \frac12 \ln \left( \frac{\hat{s}}{\muF^2}\right) \right] + {\cal R}_{qg} \,,
\label{qqqg}
\ee
where the LO Altarelli-Parisi splitting functions are
\be
P_{gg} (z) ~=~2\,  C_A\,\left[ {\cal D}_0(z) +\frac1z -2 + z(1-z) \right]
\label{Pgg} \, ,~~~~~~~~~~~
P_{gq} (z) ~=~  C_F \,\frac{1 + (1-z)^2}z~. 
\ee
In the equations above $C_A =N_c$ and $C_F = (N_c^2-1)/(\,2\,N_c)$
($N_c$ being the number of colors), $\beta_0 = (11\, C_A - 2\, N_f)/6
$ ($N_f$ being the number of active flavors) is the one-loop
$\beta$-function of the strong coupling in the SM, and
\be
{\cal D}_i (z) =  \left[ \frac{\ln^i (1-z)}{1-z} \right]_+  \label {Dfun} \, .
\ee
The two-loop virtual contributions to $g g \rightarrow A$, regularized
by the infrared-singular part of the contributions from real gluon
emission in the one-loop gluon fusion channel, $ gg \to A g$, are
displayed in the first line of eq.~(\ref{ggg}). The second line of
that equation contains the non-singular contributions from real gluon
emission. Eq.~(\ref{qqqg}) contains the contributions due to the
one-loop quark-antiquark annihilation channel, $ q \bar q \to A g $,
and to the one-loop quark-gluon scattering channel, $gq \rightarrow
qA$. General expressions for the functions ${\cal R}_{gg},\, {\cal
  R}_{q \bar q},\, {\cal R}_{q g}$ in the case of pseudoscalar
production are collected in the Appendix.

The two-loop form factor $\HA^{2\ell}$ receives contributions from
diagrams involving quarks and gluons, as well as from diagrams
involving quarks, squarks and gluinos.  The contributions from
two-loop diagrams with quarks and gluons were first computed in
ref.~\cite{SDGZ}, and later confirmed in ref.~\cite{ABDV}.  The
contribution to $\HA^{2\ell}$ arising from top-stop-gluino diagrams
was computed in ref.~\cite{HH} in the limit of vanishing pseudoscalar
mass. For what concerns the contribution arising from
bottom-sbottom-gluino diagrams, no genuine two-loop calculation has
been available so far. In the following sections we present our
calculation of both kinds of quark-squark-gluino contributions.

%%%%%%%%%%%%%%%%%%%%%%%%%%%%%%%%%%%%%%%%%%%%%%%%%%%%%%%%%%%%%%%%%%%%%%%%

\section{Technical aspects of the calculation}
\label{sec:calc}

In our computation of $\HA^{2\ell}$ we regularized the loop integrals
using the PVREG method. For the purposes of this computation, the main
advantage of PVREG is the fact that all the Lorentz indices remain
strictly 4-dimensional, thus the $\gamma_5$ matrices anticommute with
the other gamma matrices and the trace on a string of gamma matrices
can be taken using the standard 4-dimensional relations.  We recall
that in PVREG, given an ultraviolet (UV) divergent integral $I(q,m^2)$
where $q$ and $m^2$ denote collectively the external momenta and
masses, its regularized version is constructed as
\be
I^R (q,m^2, c_i, m_i^2) ~=~ I(q,m^2) ~+~ \sum_{i=1}^n \,c_i\, I(q,m_i^2)~.
\label{pvreg}
\ee
In the equation above the original integral $I(q,m^2)$ is combined
with a number $n$ of replicas, weighted by coefficients $c_i$, in
which some of the masses of the original integral are replaced by the
PV mass regulators ($m_i$), in such a way that the regularized
integral is finite if $m_i$ are kept finite, but tends to infinity as
$m_i \to \infty$. The number of added terms, as well as the relation
that the coefficients $c_i$ should satisfy in order to make $I^R$
convergent, depend on the divergent nature of the original
integral. If the latter is only logarithmically divergent, a single
subtraction is sufficient to construct $I^R$, i.e., $n=1,\, c_1 = -1,\,
m_1= M_{PV}$.  For what concerns the infrared (IR) divergences
associated to massless particles, in PVREG they are regularized by
giving a fictitious mass $\lambda$ to the massless particle, and later
considering the limit $\lambda \to 0$.

All the diagrams contributing to the virtual NLO contributions to
pseudoscalar production are at most logarithmically UV-divergent,
therefore a single subtraction is sufficient to make them
convergent. In this case, PVREG reduces to subtracting from the
original diagrams the same diagrams with some of the masses replaced
by $M_{PV}$, and then taking the limit $M_{PV} \to \infty$.  In the
case of the top-gluon contributions also the limit $\lambda \to 0$
must be taken on the fictitious gluon mass.  In the present
calculation, taking the relevant limits for the mass regulators does
not introduce additional complications with respect to the same
calculation performed in DRED or DREG.  This is due to the fact that
we are computing the two-loop diagrams via an asymptotic expansion, so
that the final result is expressed in terms of two-loop vacuum
integrals with different masses and of one-loop integrals.  Both kinds
of terms are fully known analytically, including all the relevant
limits when one or more masses are sent to infinity or to zero. The
asymptotic expansion of the relevant diagrams is generated following
the procedure described in ref.~\cite{noibot}, which amounts to adding
to and subtracting from each diagram its IR-divergent part.  As
discussed in that paper, a diagram minus its IR-divergent part can be
evaluated via a Taylor expansion in the external momenta (being this
combination IR finite by construction) while its remaining
IR-divergent part, which is expressed as a product of two one-loop
integrals, must be evaluated exactly.

In order to test our implementation of PVREG we first considered the
two-loop top-gluon contributions.  These contributions can be split in
two parts, one proportional to $C_F$ and the other proportional to
$C_A$. The latter, which stems from the non-abelian nature of $SU(3)$,
is not IR finite but contains a soft and collinear divergence that
factorizes on the lowest-order cross-section. In DREG, this IR
divergence appears as a $1/\epsilon^2$ pole multiplying the top
contribution to $\sigma^{(0)}$. We computed the top-gluon
contributions via an asymptotic expansion in the top mass up to and
including terms ${\cal O}(\ma^8/\mt^8)$.  The IR divergences are
regularized by giving a mass $\lambda$ to the gluon, while the UV
divergences are regularized by subtracting to any term a replica in
which $\lambda$ is replaced by $M_{PV}$.  The final result is then
obtained taking the limits $M_{PV} \to \infty$ and $\lambda \to 0$.
We were able to reproduce in PVREG the known result for the top-gluon
contributions obtained in DREG \cite{SDGZ,ABDV} once the PVREG
IR-divergent term $1/2\,\log^2(-\ma^2/\lambda^2)$ is identified in
DREG with $1/\epsilon^2$. This is quite non-trivial, because it is
known that, in general, regularizing the IR divergences via a
fictitious gluon mass does not respect the non-abelian symmetry of
$SU(3)$.  Thus, one expects to get the correct result only for the
part proportional to $C_F$. However, we quantize the Lagrangian
employing the Background Field Method (BFM) \cite{BFM}, so that the
external background gluons satisfy QED-like Ward identities. Then it
is not surprising that PVREG gives the correct results also for the
$C_A$ part. We also remark that within the BFM the renormalization of
the strong gauge coupling is due only to the wave function
renormalization of the external background gluons. Thus, the
renormalization of $\alpha_s$ decouples completely from the rest of
the calculation, and can be treated separately in the standard way. As
a consequence, even if PVREG is used to regularize the loop integrals,
the LO partonic cross section $\sigma^{(0)}$ can be directly expressed
in terms of the running coupling $\alpha_s (\muR)$ as in
eq.~(\ref{ggh}).

In the evaluation of the top-stop-gluino contributions to
$\HA^{2\ell}$, the two-loop integrals are regularized by subtracting
from each of them the same expression with $\tu$ and $\td$ replaced by
$M_{PV}^2$. The top-stop-gluino contributions are then computed in two
alternative ways: either by means of a Taylor expansion in the
external momentum, retaining terms of ${\cal O}(\ma^2/\mt^2)$ and
${\cal O}(\ma^2/M^2)$, or by means of an asymptotic expansion in the
superparticle masses, retaining terms up to ${\cal O}(\ma^2/M^2)$ and
${\cal O}(\mt^2/M^2)$.
The bottom-sbottom-gluino contributions to $\HA^{2\ell}$ can then be
recovered from the top-stop-gluino contributions computed with the
asymptotic expansion, by performing appropriate replacements and
taking the limit $\mb \ll \ma$. Considering the hierarchy between
$\mb$ and the other masses, we retain only terms up to ${\cal
  O}(\mb^2/\ma^2)$ and ${\cal O}(\mb/M)$.

We conclude this section with a couple of observations concerning the
use of PVREG in the computation of the virtual NLO
contributions. First, we recall that in PVREG one obtains directly the
correct result without the need of introducing a finite
renormalization factor to restore the Ward identities. Second, we note
that in PVREG the evaluation of the leading term in the Taylor
expansion (i.e., the term corresponding to $\ma=0$) does not require
the computation of counterterm diagrams. This seems natural, because
the leading term in the one-loop expression, eq.~(\ref{K1lim1}), does
not depend on the top mass. However, the same evaluation in DREG or
DRED does require the computation of counterterm diagrams. Indeed, in
$n_d$ dimensions the one-loop leading term in the Taylor expansion
contains an ${\cal O}(\epsilon)$ part that depends on the top mass, so
that the counterterm diagrams give rise to a non-vanishing
contribution.

%%%%%%%%%%%%%%%%%%%%%%%%%%%%%%%%%%%%%%%%%%%%%%%%%%%%%%%%%%%%%%%%%%%%%%%   

\section{Two-loop contributions to the form factor {\boldmath $\HA$}}
\label{sec:2loopres}

To fix our notation, we write down the Lagrangian for the interactions
of the MSSM pseudoscalar $A$ with quarks and squarks:~\footnote{Here
  and thereafter we use the notation $s_\varphi \equiv \sin\varphi, \,
  c_\varphi \equiv \cos\varphi$ for a generic angle $\varphi$.}
\be
\label{couplings}
{\cal L} ~\supset~
\frac{i}{\sq2}\,h_t\,\cbe~ A\,\bar t \gamma_5 t \,+\,
\frac{i}{\sq2}\,h_b\,\sbe~ A\,\bar b \gamma_5 b \,+\,
\frac{i}{\sq2}\,\left(
h_t\,\cbe\,Y_t\,A\,\stu^*\std\,+\, 
h_b\,\sbe\,Y_b\,A\,\sbu^*\sbd~~-~{\rm h.c.}\,\right) ~,
\ee
where: $h_t$ and $h_b$ are the top and bottom Yukawa couplings; $Y_t =
A_t - \mu\tan\beta$ and $Y_b = A_b - \mu\cot\beta$; $A_t$ and $A_b$
are the soft SUSY-breaking Higgs-squark-squark couplings; $\mu$ is the
Higgs mass term in the MSSM superpotential. Our convention for the
sign of $\mu$ is such that, e.g., the stop and sbottom left-right
mixing angles $\theta_t$ and $\theta_b$ obey the relations
\be
\sdt~=~\frac{2\,\mt\,(A_t+\mu\cot\beta)}{\tu-\td}~,~~~~~~~~
\sdb~=~\frac{2\,\mb\,(A_b+\mu\tan\beta)}{\bu-\bd}~.
\label{convmu}
\ee

\begin{figure}[t]
\begin{center}
\mbox{
\epsfig{figure=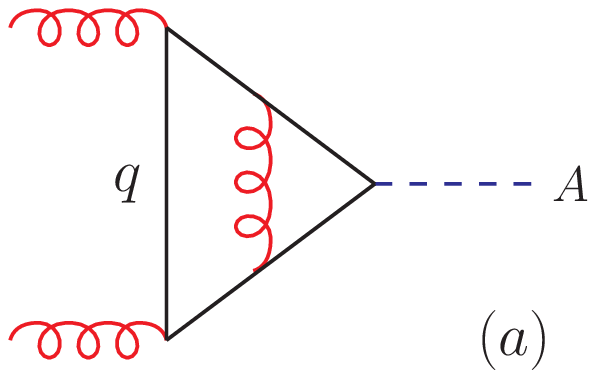,width=4.9cm}~~~~~~~
\epsfig{figure=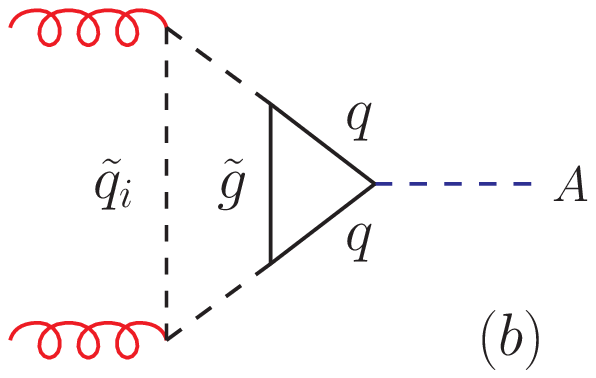,width=4.9cm}~~~~~~~
\epsfig{figure=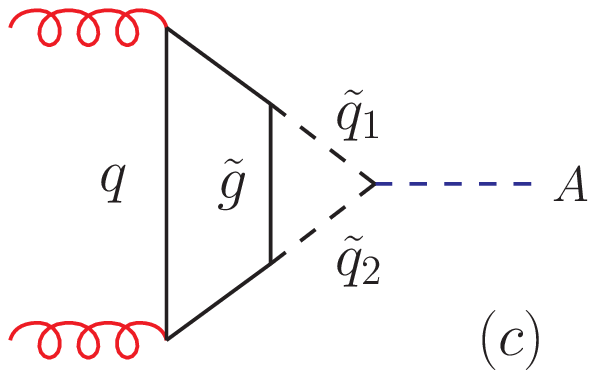,width=4.9cm}
}
\caption{Examples of two-loop quark-gluon diagrams $(a)$, and of
  two-loop quark-squark-gluino diagrams involving $(b)$ the
  pseudoscalar-quark coupling or $(c)$ the pseudoscalar-squark
  coupling. Here, $q = t, b$ and $i=1,2$.}
\label{fig:diags}
\end{center}
\end{figure}

The fact that the pseudoscalar only couples to two different squark
mass eigenstates, while gluons only couple to two equal eigenstates,
implies that the form factor $\HA$ receives neither one-loop
contributions from diagrams with squarks nor two-loop contributions
from diagrams with squarks and gluons. However, contributions to
$\HA^{2\ell}$ do arise from two-loop diagrams with quarks and gluons,
as well as from two-loop diagrams with quarks, squarks and
gluinos. Examples of such diagrams, involving either the
pseudoscalar-quark coupling or the pseudoscalar-squark coupling, are
given in figure \ref{fig:diags}.

The two-loop form factor for pseudoscalar production can be decomposed
as
\be
\HA^{2\ell}~=~ T_F\,\left[
\cot\beta\,\left({\cal K}_{tg}^{2\ell}\,+ \, \HAt^{2\ell}\right)
\,+\, \tan\beta\,\left( {\cal K}_{bg}^{2\ell}\, +\, \HAb^{2\ell}\right)
\right]~,
\label{HA2l}
\ee
where ${\cal K}_{qg}^{2\ell}$ denotes the quark-gluon contributions
($q=t,b$), and ${\cal K}_{q\tilde q\tilde g}^{2\ell}$ denotes the
quark-squark-gluino contributions. In the following we discuss
separately the two-loop contributions arising from quark-gluon,
top-stop-gluino and bottom-sbottom-gluino diagrams.

\subsection{Quark-gluon contributions}
\label{sec:quarkglu}

We recall for completeness the results of refs.~\cite{SDGZ,ABDV} for
the contributions to $\HA^{2\ell}$ arising from diagrams with quarks
and gluons (see figure \ref{fig:diags}a). If the corresponding
contribution in the one-loop form factor $\HA^{1\ell}$ is expressed in
terms of the physical quark mass, the two-loop contribution for a
given quark $q$ reads
\be
{\cal K}_{qg}^{2\ell} = C_F\,\left[{\cal F}_1(\tau_q) + 
\frac43\,{\cal F}_2(\tau_q)\right] + C_A\,{\cal F}_3(\tau_q)~,
\label{K2lonshell}
\ee
If the one-loop form factor is instead expressed in terms of the
running quark mass, renormalized in the $\drbar$ scheme at the scale
$Q$, the two-loop contribution becomes
\be
{\cal K}_{qg}^{2\ell} = C_F\,\left[{\cal F}_1(\tau_q) + 
{\cal F}_2(\tau_q)\left(\ln\frac{m_q^2}{Q^2} - \frac13 \right)\right] 
+ C_A\,{\cal F}_3(\tau_q)~.
\label{K2ldrbar}
\ee
Expressions for the functions denoted here as ${\cal F}_1(\tau)$,
${\cal F}_2(\tau)$ and ${\cal F}_3(\tau)$, valid for arbitrary values
of $\tau$, can be found in ref.~\cite{ABDV}. They correspond to the
functions ${\cal E}_t^{\,(2\ell,a)}(4/\tau)$ in eq.~(4.6), ${\cal
  E}_t^{\,(2\ell,b)}(4/\tau)$ in eq.~(4.7), and ${\cal
  K}_t^{\,(2\ell,C_A)}(4/\tau)$ in eq.~(4.12) of that paper,
respectively. Their limiting behaviors for heavy and light quark are
\bea
(\tau \gg 1):~~ &{\cal F}_1(\tau)& 
\longrightarrow~ -\frac{4}{3\tau} ~+~ {\cal O}(\tau^{-2})~,\\
&{\cal F}_2(\tau)& 
\longrightarrow~ -\frac{1}{\tau} ~+~ {\cal O}(\tau^{-2})~,\\
&{\cal F}_3(\tau)& 
\longrightarrow~ -2 -\frac{1}{6\tau} ~+~ {\cal O}(\tau^{-2})~,\\\nn\\
(\tau \ll 1):~~ &{\cal F}_1(\tau)&
\longrightarrow~ -\tau\,\biggr[\frac95 \,\zeta_2^2 
- \zeta_3+(2-\zeta_2-4\,\zeta_3)\,
\ln(\frac{-4}{\tau}) \nn\\
&&~~~~~~~~~~~~~ - (1-\zeta_2)\ln^2(\frac{-4}{\tau})
+\frac14\,\ln^3(\frac{-4}{\tau})+\frac1{48}\,\ln^4(\frac{-4}{\tau})\biggr]
 ~+~ {\cal O}(\tau^{2})~,\\\nn\\
&{\cal F}_2(\tau)& 
\longrightarrow~ \frac{3\,\tau}4\,\biggr[2\,\ln(\frac{-4}{\tau})
-\ln^2(\frac{-4}{\tau})\biggr] ~+~ {\cal O}(\tau^{2})~,\\\nn\\
 &{\cal F}_3(\tau)&
\longrightarrow~ \tau\,\biggr[\frac85 \,\zeta_2^2 
+ 3\,\zeta_3- 3\,\zeta_3\,\ln(\frac{-4}{\tau}) 
+ \frac14\,(1+2\,\zeta_2)\ln^2(\frac{-4}{\tau}) \nn\\
&&~~~~~~~~~~~
+\frac1{48}\,\ln^4(\frac{-4}{\tau})\biggr]
 ~+~ {\cal O}(\tau^{2})~.
\eea

\subsection{Top-stop-gluino contributions}
\label{sec:topstop}

While a fully analytic computation of the top-stop-gluino
contributions to $\HA^{2\ell}$ valid for arbitrary values of all the
relevant particle masses is currently beyond our reach, it is possible
to derive approximate analytic results valid in different
phenomenologically relevant limits.

To start with, we computed the term $\HAt^{2\ell}$ in eq.~(\ref{HA2l})
via a Taylor expansion in the external Higgs momentum up to terms of
${\cal O}(\ma^2/\mt^2)$ and ${\cal O}(\ma^2/M^2)$, where $M$ denotes
generically the stop and gluino masses. Such expansion should give a
reasonable approximation to the full result when $\ma$ is small
compared to the other masses, and is anyway restricted to values of
$\ma$ below the lowest threshold encountered in the diagrams (this
usually means $\ma<2\,\mt$). In the limit of vanishing $\ma$ we find
that our result for $\HAt^{2\ell}$ can be cast in an extremely compact
form:
\be
\HAt^{2\ell}~=~
\left(\frac{\sdt}2 - \frac{\mt\,Y_t}{\tu-\td}\right)\,
\left[f(\g,\t,\tu)-f(\g,\t,\td)\right]~,
\label{compact}
\ee
where
\bea
f(\g,\t,\ti)&=&
C_F\,\frac{\mg}{\mt\,\Delta}\,\biggr[
\t\,(\g-\t+\ti)\,\ln\frac{\t}{\g} + \ti\,(\g+\t-\ti)\,\ln\frac{\ti}{\g}\nn\\
&&~~~~~~~~~~~~~~~+2\,\g\,\t\,\Phi(\g,\t,\ti)\biggr]\nn\\
&+&C_A\,\frac{\mt}{\mg\,\Delta}\,\biggr[
\ti\,(\ti-\t-\g)\,\ln\frac{\t}{\g} + \ti\,(\t-\ti-\g)\,\ln\frac{\ti}{\g}\nn\\
&&~~~~~~~~~~~~~~~+\g\,(\t+\ti-\g)\,\Phi(\g,\t,\ti)\biggr]~,
\label{ffunc}
\eea
the function $\Phi(\g,\t,\ti)$ is given, e.g., in appendix A of
ref.~\cite{dedeslav}, and we introduced the shortcut $\Delta = \mt^4 +
\gq + \tiq - 2\,(\t\,\g + \t\, \ti + \g\, \ti)$\,. As appears from
eqs.~(\ref{H1lp}) and (\ref{K1lim1}), in the limit of vanishing $\ma$
the one-loop top contribution to $\HA$ reduces to $-\cot\beta$, i.e.,
it does not actually depend on any parameter subject to ${\cal
  O}(\as)$ corrections. Therefore, the results in eqs.~(\ref{compact})
and (\ref{ffunc}) do not depend on the renormalization scheme in which
the calculation is performed. The contributions to $\HAt^{2\ell}$ of
the first order in the Taylor expansion in $\ma^2$ are too lengthy to
be printed here, but in section \ref{sec:num} we will discuss their
relevance in a representative region of the MSSM parameter space.

The two terms between parentheses in eq.~(\ref{compact}) come from the
diagrams with pseudoscalar-top and pseudoscalar-stop couplings in
figures \ref{fig:diags}b and \ref{fig:diags}c, respectively.
Inserting the explicit expressions for $\sdt$ and $Y_t$ we find
\be
\HAt^{2\ell} ~=~
\frac{\mt\,\mu}{\tu-\td}\,(\cot\beta+\tan\beta)\,
\left[f(\g,\t,\tu)-f(\g,\t,\td)\right]~,
\label{compact2}
\ee
i.e., the explicit dependence of $\HAt^{2\ell}$ on $A_t$ drops out,
leaving only a dependence on $\mu$. Ref.~\cite{HH} points out that
this happens because the $\mu$ term breaks the axial U(1) Peccei-Quinn
symmetry of the MSSM potential, thus violating the Adler-Bardeen
theorem \cite{adler} which would otherwise guarantee the cancellation
of all contributions from irreducible diagrams beyond one loop.

We compared our result for $\HAt^{2\ell}$ in the limit of vanishing
$\ma$, eqs.~(\ref{compact})--(\ref{ffunc}), with the result for the
coefficient $\tilde c_1^{\,(1)}$ defined in ref.~\cite{HH}. That
result was deemed too voluminous to be printed explicitly in
ref.~\cite{HH}, and was made available in the fortran code {\tt
  evalcsusy.f} \cite{HS}. We find full numerical agreement with {\tt
  evalcsusy.f}, after taking into account that $\tilde c_1^{\,(1)} =
-T_F\,\cot\beta\,\HAt^{2\ell}$ and that ref.~\cite{HH} employs the
opposite convention for the sign of $\mu$ with respect to our
eq.~(\ref{convmu}).

Even when the superparticles are much heavier than the pseudoscalar,
the validity of the result for $\HAt^{2\ell}$ obtained via a Taylor
expansion in $\ma^2$ becomes questionable if $\ma$ is close to or even
larger than $\mt$. To cover this region of the parameter space we
performed an asymptotic expansion of $\HAt^{2\ell}$ in the large
superparticle masses. More specifically, we consider the case
$(\ma,\mt) \ll M$ without assuming any hierarchy between $\ma$ and
$\mt$, and retain terms up to ${\cal O}(\ma^2/M^2)$ and ${\cal
  O}(\mt^2/M^2)$ in the expansion. Assuming that the top contribution
to $\HA^{1\ell}$ in eqs.~(\ref{H1lp}) and (\ref{K1l}) is expressed in
terms of the pole top mass, we find
\bea
\HAt^{2\ell} &=& -\frac{C_F}2\,{\cal K}^{1\ell}(\tau_t)\, \frac{\mg}{\mt}
\,\biggr(\frac{\sdt}{2} - \frac{\mt\,Y_t}{\tu-\td}\biggr)
\left(\frac{x_1}{1-x_1} \ln x_1-\frac{x_2}{1-x_2}\ln x_2\right)\nn\\\nn\\
&& -~ \frac{\mt}{\mg}\,\,\sdt\,{\cal R}_1 
~+~\frac{2\,\mt^2\,Y_t}{\mg\,(\tu-\td)}\,{\cal R}_2 
~+~ \frac{\mt^2}{\mg^2}\,{\cal R}_3
~-~\frac12\,{\cal K}^{1\ell}(\tau_t)\,\frac{\ma^2}{\tu-\td}\,{\cal R}_4~,
\label{Kttg2}
\eea
where $x_i = \ti/\g\,$, the one-loop function ${\cal K}^{1\ell}(\tau)$
was defined in eq.~(\ref{K1l}), and the terms ${\cal R}_i$ collect
contributions suppressed by $\mt/M$ or $\mt^2/M^2$:
\bea
\label{R1}
{\cal R}_1 &=&
\frac{C_F}{4\,(1-x_1)^3}\,\biggr[(1-x_1^2+2\,x_1\,\ln x_1)
\biggr(2\,\ln\frac{\mg^2}{\mt^2} 
- 3 - \frac32 {\cal K}^{1\ell}(\tau_t) + 2\,{\cal B}\biggr)\nn\\
&&~~~~~~~~~~~~~~~~~
-8\,x_1\,{\rm Li_2}(1-x_1) -2\,x_1\,(3+x_1)\,\ln x_1\biggr]\nn\\
&+&\frac{C_A}{2\,(1-x_1)^2}\,\biggr[(1-x_1+x_1\,\ln x_1)
\biggr(\ln\frac{\mt^2}{\mg^2} + 1
+\frac12 {\cal K}^{1\ell}(\tau_t)-{\cal B}\biggr)\nn\\
&&~~~~~~~~~~~~~~~~~
+2\,x_1\,{\rm Li_2}(1-x_1) +x_1\,(1+x_1)\,\ln x_1\biggr]\nn\\\nn\\
&+&\frac{C_F}{(x_1-x_2)^2}\,\frac{Y_t}{\mg}\,
\left(1+ \frac12 {\cal K}^{1\ell}(\tau_t)\right)\biggr[
\frac{x_1^2\,(1-2\,x_2)}{2(1-x_1)(1-x_2)} 
+\frac{x_1}{2(1-x_1)^2}\,(x_1^2-2\,x_2+x_1\,x_2)\,\ln x_1 \biggr]\nn\\\nn\\
&-& \biggr(x_1 \longleftrightarrow x_2\biggr)~,\\\nn\\
{\cal R}_2 &=&
\frac{C_F}{4\,(1-x_1)^3}\,\biggr[2\,(1-x_1^2+2\,x_1\,\ln x_1)\,
\ln\frac{\mg^2}{\mt^2}-8\,x_1\,{\rm Li_2}(1-x_1)\nn\\
&&~~~~~~~~~~~~~~~~~ 
+(1-x_1^2)\,\biggr(1+\frac12 {\cal K}^{1\ell}(\tau_t)\biggr)
-2\,x_1\biggr(2+x_1-\frac12 {\cal K}^{1\ell}(\tau_t)\biggr) \ln x_1\biggr]\nn\\
&+&\frac{C_A}{2\,(1-x_1)^2}\,\biggr[(1-x_1+x_1\,\ln x_1)\,
\ln\frac{\mt^2}{\mg^2}+2\,x_1\,{\rm Li_2}(1-x_1) +x_1\,(1+x_1)\,
\ln x_1\biggr]\nn\\\nn\\
&-& \biggr(x_1 \longleftrightarrow x_2\biggr)~,\\\nn\\
{\cal R}_3 &=&
\frac{C_F}{6\,(1-x_1)^4}\,(-2-3\,x_1+6\,x_1^2-x_1^3-6\,x_1\,\ln x_1)
\biggr(2 +{\cal K}^{1\ell}(\tau_t)- {\cal B}\biggr)\nn\\
&+&\frac{C_A}{8\,(1-x_1)^3}\,(1-x_1^2+2\,x_1\,\ln x_1)\,
\biggr(2 +{\cal K}^{1\ell}(\tau_t)- 2\,{\cal B}\biggr)
~+~\biggr(x_1 \longleftrightarrow x_2\biggr)~,\\\nn\\
\label{R4}
{\cal R}_4 &=&
\frac{C_F}{(x_1-x_2)^2}\,\frac{Y_t}{\mg}\,\biggr[
\frac{x_1^2\,(1-2\,x_2)}{2(1-x_1)(1-x_2)} 
+\frac{x_1}{2(1-x_1)^2}\,(x_1^2-2\,x_2+x_1\,x_2)\,\ln x_1 \biggr]
~-~ \biggr(x_1 \longleftrightarrow x_2\biggr)~.\nn\\
\eea
In the equations above, ${\cal B}$ denotes the finite part of the
Passarino-Veltman function $B_0(\ma^2,\mt^2,\mt^2)$ computed at the
renormalization scale $Q^2 = \mt^2$. The comparison between the result
for $\HAt^{2\ell}$ obtained via a Taylor expansion in $\ma^2$ and the
corresponding result obtained via an asymptotic expansion in $M$ will
be discussed in section \ref{sec:num}.

\subsection{Bottom-sbottom-gluino contributions}
\label{sec:botsbot}

A result for the bottom-sbottom-gluino contribution $\HAb^{2\ell}$ can
be obtained by performing the obvious replacement $t\rightarrow b$ in
the result for $\HAt^{2\ell}$ obtained via the asymptotic expansion in
$M$, eqs.~(\ref{Kttg2})--(\ref{R4}). Considering that $\mb\ll\ma$, and
that we are assuming $\ma\ll M$, we retain only the terms up to ${\cal
  O}(\mb/M)$ and ${\cal O}(\mb^2/\ma^2)$. In particular, the terms
${\cal R}_2$, ${\cal R}_3$ and ${\cal R}_4$ in eq.~(\ref{Kttg2})  give
contributions of higher order in $\mb$ and can be neglected, while in
the expression for ${\cal R}_1$, eq.~(\ref{R1}), we drop the
occurrences of ${\cal K}^{1\ell}(\tau_b)$ and use ${\cal B} = 2 -
\ln(-\ma^2/\mb^2)$.  As a result, assuming that the bottom
contribution to $\HA^{1\ell}$ in eqs.~(\ref{H1lp}) and (\ref{K1lim2})
is fully expressed in terms of the pole bottom mass, we again find a
rather compact expression for the term $\HAb^{2\ell}$ in
eq.~(\ref{HA2l}):
\be
\HAb^{2\ell} ~=\, -\frac{C_F}2\,{\cal K}^{1\ell}(\tau_b)\, \frac{\mg}{\mb}
\,\biggr(\frac{\sdb}{2} - \frac{\mb\,Y_b}{\bu-\bd}\biggr)
\left(\frac{x_1}{1-x_1} \ln x_1-\frac{x_2}{1-x_2}\ln x_2\right)
- \frac{\mb}{\mg}\,\,\sdb\,{\cal R}_1~.
\label{bsbg1}
\ee
Here $x_i = \bi/\g\,$, and ${\cal R}_1$ collects the
contributions suppressed by $\mb/M$:
\bea
{\cal R}_1 &=&
\frac{C_F}{4\,(1-x_1)^3}\,\biggr[(1-x_1^2+2\,x_1\,\ln x_1)\biggr(1
-2\,\ln(\frac{-\ma^2}{~\g})\biggr)
-8\,x_1\,{\rm Li_2}(1-x_1) -2\,x_1\,(3+x_1)\,\ln x_1\biggr]\nn\\
&+&\frac{C_A}{2\,(1-x_1)^2}\,\biggr[(1-x_1+x_1\,\ln x_1)
\biggr(\ln(\frac{-\ma^2}{~\g})-1\biggr)
+2\,x_1\,{\rm Li_2}(1-x_1) +x_1\,(1+x_1)\,\ln x_1\biggr]\nn\\\nn\\
&+&\frac{C_F}{(x_1-x_2)^2}\,\frac{Y_b}{\mg}\,\biggr[
\frac{x_1^2\,(1-2\,x_2)}{2(1-x_1)(1-x_2)} 
+\frac{x_1}{2(1-x_1)^2}\,(x_1^2-2\,x_2+x_1\,x_2)\,\ln x_1 \biggr]\nn\\\nn\\
&-& \biggr(x_1 \longleftrightarrow x_2\biggr)~.
\eea
As in the case of the top-stop-gluino contribution, the terms
proportional to $Y_b$ originate from the diagrams that involve the
pseudoscalar-sbottom coupling, while the other terms originate from
the diagrams that involve the pseudoscalar-bottom coupling. Inserting
the expressions for $\sdb$ and $Y_b$ in the first term in the
right-hand side of eq.~(\ref{bsbg1}) we obtain
\be
\HAb^{2\ell} ~=\, -\frac{C_F}2\,{\cal K}^{1\ell}(\tau_b)\, 
\frac{\mg\, \mu}{\bu-\bd}\,(\tan\beta+\cot\beta)\,
\left(\frac{x_1}{1-x_1} \ln x_1-\frac{x_2}{1-x_2}\ln x_2\right)
- \frac{\mb}{\mg}\,\,\sdb\,{\cal R}_1~.
\label{bsbg2}
\ee

Similarly to what found in ref.~\cite{noibot} for the production of
CP-even Higgs bosons, if the one-loop contribution to $\HA$ is
expressed in terms of the pole bottom mass the bottom-sbottom-gluino
diagrams induce potentially large two-loop contributions. According to
whether or not we insert the explicit expression for $\sdb$ in our
formulae, such contributions manifest themselves either as terms
enhanced by the ratio $\mg/\mb$, as in eq.~(\ref{bsbg1}), or as terms
enhanced by $\tan\beta$, as in eq.~(\ref{bsbg2}). However, such terms
cancel out if the pseudoscalar-bottom coupling entering the one-loop
contribution to $\HA$ is identified with the $\drbar$-renormalized
mass $\widehat \mb$, while the mass of the bottom quark running in the
loop is identified with the pole mass $M_b$ (this amounts to rescaling
by $\widehat \mb/M_b$ the one-loop result fully computed in terms of
$M_b$). As a result, the two-loop form factor in eq.~(\ref{HA2l}) is
shifted as
\be
\HA^{2\ell} ~\longrightarrow~\HA^{2\ell}  ~-~
\tan\beta\,{\mathcal K}^{1\ell} (\tau_b)\, T_F\,C_F
\left[ \frac34\,\ln \frac{\mb^2}{Q^2} -\frac54 +
\dmbsusy\right]
\label{shift}
\ee
with respect to the result obtained when the one-loop bottom
contribution is fully expressed in terms of $M_b$. Here $Q$ is the
scale at which the running mass $\widehat \mb$ is renormalized, and
$(\delta m_b)^{\scriptscriptstyle SUSY}$ denotes the SUSY contribution
to the bottom self-energy, in units of $C_F\,\alpha_s/\pi$ and in the
limit of vanishing $m_b\,$:
\be
\label{dmbovmb}
\dmbsusy ~=~ -\frac14\,\left[
\ln\frac{\mg^2}{Q^2}  + f(x_1)+f(x_2) + \frac{\mg}{\mb}\,\sdb\,
\left(\frac{x_1}{1-x_1} \ln x_1-\frac{x_2}{1-x_2}\ln x_2\right)\right]~,
\ee
where 
\be
f(x) ~=~ \frac{ x - 3}{4\, (1-x)} + \frac{x\,(x-2)}{2\,(1-x)^2} \,\ln x ~.
\ee

While the shift in eq.~(\ref{shift}) removes the contributions
enhanced by $\mg/\mb$ (or $\tan\beta$), it does introduce potentially
large logarithms of the ratio between the renormalization scale $Q$
and the masses of the particles running in the loop. Such logarithms
cannot be eliminated by a specific scale choice for $\widehat \mb$,
unless $Q$ is set to a value much smaller than the bottom mass
itself. Therefore, as already found in ref.~\cite{noibot} for the
CP-even Higgs bosons, the bottom contributions to $\HA^{2\ell}$ may
turn out to be sizable even in the ``mixed'' renormalization scheme in
which the $\tan\beta$-enhanced contributions are absorbed in a
redefinition of the pseudoscalar-bottom coupling entering
$\HA^{1\ell}$.

Finally, if the bottom contribution to $\HA^{1\ell}$ is fully
expressed in terms of the running bottom mass $\widehat \mb$ the
bottom-sbottom-gluino contribution to the form factor in
eq.~(\ref{bsbg1}) is shifted as
\be
\HAb^{2\ell} ~\longrightarrow~ \HAb^{2\ell} 
~+~ \frac43 \,C_F\,{\cal F}_2(\tau_b)\,\dmbsusy~.
\label{shiftDR}
\ee
In this case $\HA^{2\ell}$ contains both terms enhanced by $\mg/\mb$
and potentially large logarithms, the latter arising from $(\delta
m_b)^{\scriptscriptstyle SUSY}$ in eq.~(\ref{shiftDR}) as well as from
the two-loop bottom-gluon contribution in eq.~(\ref{K2ldrbar}).

\subsection{Comparison with the effective-Lagrangian approximation}
\label{sec:efflag}

It is well known that, in the MSSM, loop diagrams involving
superparticles induce interactions between the quarks and the
``wrong'' Higgs doublets, i.e., interactions that are absent from the
tree-level Lagrangian due to the requirement that the superpotential
be a holomorphic function of the superfields~\cite{hrs}. Such
non-holomorphic, loop-induced Higgs-quark interactions result in
$\tan\beta$-enhanced (or $\tan\beta$-suppressed) corrections to the
MSSM predictions for various physical observables.  If all
superparticles are considerably heavier than the Higgs bosons they can
be integrated out of the Lagrangian, in which case the loop-induced
corrections are {\em resummed} in effective Higgs-quark couplings. In
particular, if $g_b^{\phi}$ denote the tree-level couplings of a
neutral Higgs $\phi = (h,H,A)$ to bottom quarks (normalized to the SM
value), the corresponding effective couplings $\tilde g_b^{\phi}$
read~\cite{effL,GHS}
\be
\label{effcoup}
\tilde g_b^{h} ~=~ \frac{g_b^{h}}{1+\Delta_b}
\left(1-\Delta_b\,\frac{\cot\alpha}{\tan\beta}\right),~~~~~
\tilde g_b^{H} ~=~ \frac{g_b^{H}}{1+\Delta_b}
\left(1+\Delta_b\,\frac{\tan\alpha}{\tan\beta}\right),~~~~~
\tilde g_b^{A} ~=~ \frac{g_b^{A}}{1+\Delta_b}
\left(1-\Delta_b\,\cot^2\beta\right),
\ee
where $\alpha$ is the mixing angle in the CP-even Higgs sector and, 
to ${\cal O}(\alpha_s)$,
\be
\label{deltab}
\Delta_b ~=~ \frac{\alpha_s\,C_F}{2\pi}\,\frac{\mg\,\mu\,\tan\beta}
{\bu-\bd}\,\left(\frac{\x1g}{1-\x1g} \ln \x1g
-\frac{x_2}{1-x_2}\ln x_2\right)~.
\ee

In the calculation of processes involving the Higgs-bottom couplings,
it is often found that the $\tan\beta$-enhanced corrections can be
included to all orders in an expansion in powers of
$\alpha_s\tan\beta$ by inserting the effective couplings of
eq.~(\ref{effcoup}) in the lowest-order result. A comparison with our
explicit results for the two-loop form factors allows us to test the
validity of that procedure in the case of the production of both
CP-even~\cite{noibot} and CP-odd Higgs bosons in gluon
fusion.\footnote{A comparison for the light scalar $h$ in the limit of
  vanishing sbottom mixing was discussed in ref.~\cite{hhm}, and a
  numerical comparison for the heavy scalar $H$ was shown, without a
  detailed discussion, in ref.~\cite{spiraDb}.}

We recall that the bottom-quark contributions ${\cal
  H}^{1\ell\,,b}_\phi$ to the one-loop form factors for the production
of the Higgs boson $\phi=(h,H,A)$ read
\be
\label{decoupl}
{\cal H}^{1\ell\,,b}_h ~=~ -T_F\,\frac{\sin\alpha}{\cos\beta}
~{\cal G}_{1/2}^{1\ell}(\tau_b)~,~~~~~~
{\cal H}^{1\ell\,,b}_H ~=~ T_F\,\frac{\cos\alpha}{\cos\beta} 
~{\cal G}_{1/2}^{1\ell}(\tau_b)~,~~~~~~
{\cal H}^{1\ell\,,b}_A ~=~ T_F\, \tan\beta~{\cal K}^{1\ell}(\tau_b)~,
\ee
where the function ${\cal G}_{1/2}^{1\ell}(\tau)$ is given, e.g., in
eq.~(12) of ref.~\cite{noibot}. Assuming that ${\cal
  H}^{1\ell\,,b}_\phi$ are expressed in terms of the pole bottom mass,
and that the Higgs-sbottom couplings are renormalized in a way that
avoids the introduction of additional $\tan\beta$-enhanced corrections
(see ref.~\cite{noibot}), we find that the two-loop form factors read
\bea
\label{effHh}
{\cal H}^{2\ell}_h &=& ~~\,{\cal H}^{1\ell\,,b}_h~\left[
-\frac{\pi}{\alpha_s}\,\Delta_b\,
\left(1+\,\frac{\cot\alpha}{\tan\beta}\right) + 
\frac{C_F}{4}\,\frac{A_b-\mu\cot\alpha}{\mg}\,\sdb^2\,g(x_1,x_2)\right]
~+~\ldots~,\\
\label{effHH}
{\cal H}^{2\ell}_H &=& ~~\,{\cal H}^{1\ell\,,b}_H~\left[
-\frac{\pi}{\alpha_s}\,\Delta_b\,
\left(1-\,\frac{\tan\alpha}{\tan\beta}\right) +
\frac{C_F}{4}\,\frac{A_b+\mu\tan\alpha}{\mg}\,\sdb^2\,g(x_1,x_2)\right]
~+~\ldots~,\\
\label{effHA}
{\cal H}^{2\ell}_A &=& -{\cal H}^{1\ell\,,b}_A~~
~\frac{\pi}{\alpha_s}\,\Delta_b\,(1+\cot^2\beta) +~\ldots~,
\eea
where the ellipses denote contributions suppressed by $\mb/M$ or
$\mz^2/M^2$, as well as all of the contributions from diagrams
involving top and stop, and
\be
\label{funcg}
g(x_1,x_2) ~=~ \frac{1}{1-x_1}\left(1+\frac{\ln x_1}{1-x_1}\right)
+ \frac{1}{1-x_2}\left(1+\frac{\ln x_2}{1-x_2}\right)
-\frac{2}{x_1-x_2}\,\left(\frac{\x1g}{1-\x1g} \ln \x1g
-\frac{x_2}{1-x_2}\ln x_2\right)~.
\ee

In practice, the effective-Lagrangian approximation consists in
rescaling the one-loop bottom contributions ${\cal
  H}^{1\ell\,,b}_\phi$ by the same factors that rescale the
Higgs-bottom couplings $g_b^\phi$ in eq.~(\ref{effcoup}).  Expanding
the rescaling factors to the first order in $\Delta_b$ it is easy to
see that the effective-Lagrangian approximation does indeed reproduce
the two-loop terms proportional to $\Delta_b$ in
eqs.~(\ref{effHh})--(\ref{effHA}).

It is also interesting to consider the so-called decoupling limit of
the MSSM, $\ma\gg\mz$, in which $\cot\alpha \rightarrow -\tan\beta$
and the light scalar $h$ has SM-like couplings to fermions and gauge
bosons.\footnote{The validity of the effective-Lagrangian
  approximation for the light scalar $h$ in the decoupling limit was
  already discussed in ref.~\cite{GHS} in the context of Higgs boson
  decays to bottom quark pairs.} Eq.~(\ref{effcoup}) shows that in
this limit the effective coupling of $h$ to bottom quarks is equal to
the tree-level coupling, therefore in the effective-Lagrangian
approximation there are no $\tan\beta$-enhanced contributions to
${\cal H}^{2\ell}_h$. Indeed, for $\cot\alpha \rightarrow -\tan\beta$
the terms proportional to $\Delta_b$ drop out of the two-loop form
factor in eq.~(\ref{effHh}). However, eq.~(\ref{effHh}) also shows
that in the decoupling limit ${\cal H}^{2\ell}_h$ contains additional
$\tb$-enhanced contributions, controlled by the left-right sbottom
mixing $X_b = (A_b+\mu\tb)$, which are not reproduced by the
effective-Lagrangian approximation. However, when the implicit
dependence of the sbottom masses and mixing on the bottom mass is
taken into account, such contributions turn out to be partially
suppressed by powers of $\mb$. Indeed, taking for illustrative
purposes the limit in which the diagonal entries of the sbottom mass
matrix as well as the squared gluino mass are all equal to $M^2$, and
expanding the form factor in powers of $\mb$, we find
\be
\label{missed}
{\cal H}^{2\ell}_h ~\supset~ -{\cal H}^{1\ell\,,b}_h~\,
\frac{C_F}{12}~\frac{\mb^2\,X_b^3}{M^5}
~+~ T_F\,\frac{2\,C_A+25\,C_F}{18}\,\frac{\mb^2\, X_b^2}{M^4} 
~+~\ldots~,
\ee
where the ellipses denote terms further suppressed by powers of $\mb$
or $\mz$, as well as all of the contributions from diagrams involving
top and stop. The first term in eq.~(\ref{missed}) comes from the
expansion of the terms proportional to $\sdb^2$ in eq.~(\ref{effHh}),
while the second comes from the expansion of terms not shown in
eq.~(\ref{effHh}). The contributions neglected by the
effective-Lagrangian approximation can be relevant for values of $X_b$
large enough to compensate for the suppression due to $\mb$. It should
however be recalled that in the decoupling limit ${\cal
  H}^{1\ell\,,b}_h$ is not further enhanced by $\tan\beta$, therefore
-- differently from what happens in the case of the heavy Higgs bosons
-- the total form factor for $h$ production can still be dominated by
the top/stop contributions even for large values of $\tan\beta$.

%%%%%%%%%%%%%%%%%%%%%%%%%%%%%%%%%%%%%%%%%%%%%%%%%%%%%%%%%%%%%

\section{Numerical examples}
\label{sec:num}

We will now illustrate the effect of the two-loop quark-squark-gluino
contributions to the form factor for pseudoscalar Higgs production in
a representative region of the MSSM parameter space.

The SM parameters entering our calculation include the $Z$ boson mass
$\mz = 91.1876$ GeV, the $W$ boson mass $\mw = 80.399$ GeV and the
strong coupling constant $\alpha_s(\mz) = 0.118$ \cite{PDG}. For the
pole masses of the top and bottom quarks we take $M_t = 173.3$ GeV
\cite{topmass} and $M_b = 4.49$ GeV, the latter corresponding to the
SM running mass (in the $\msbar$ scheme) $\overline \mb(\mb) = 4.16$
GeV \cite{botmass}. 

Since the squarks do not contribute to the one-loop amplitude for
pseudoscalar production, the only parameters entering $\HA^{1\ell}$ in
addition to the quark masses are $\tan\beta$ and $\ma$. Neither of
those parameters is subject to one-loop $\oas$ corrections, therefore
we need not specify a renormalization scheme for them (although it is
natural to consider $\ma$ as the pole pseudoscalar mass). The
remaining input parameters are $\mg,\,\mu,\,A_t,\,A_b$ and the soft
SUSY-breaking mass terms for stop and sbottom squarks, $m_Q,\,m_U$ and
$m_D$. Since these parameters only enter the two-loop part of the form
factor we need not specify a renormalization scheme for them
either. For simplicity, in our numerical examples we will set all the
SUSY-breaking parameters, as well as the supersymmetric mass parameter
$\mu$, to a common value $M$. Note however that the squark mass
eigenstates will differ from $M$, because of the supersymmetric
(F-term and D-term) contributions to the squark mass matrices as well
as of the left-right mixing terms.

\begin{figure}[p]
\begin{center}
\epsfig{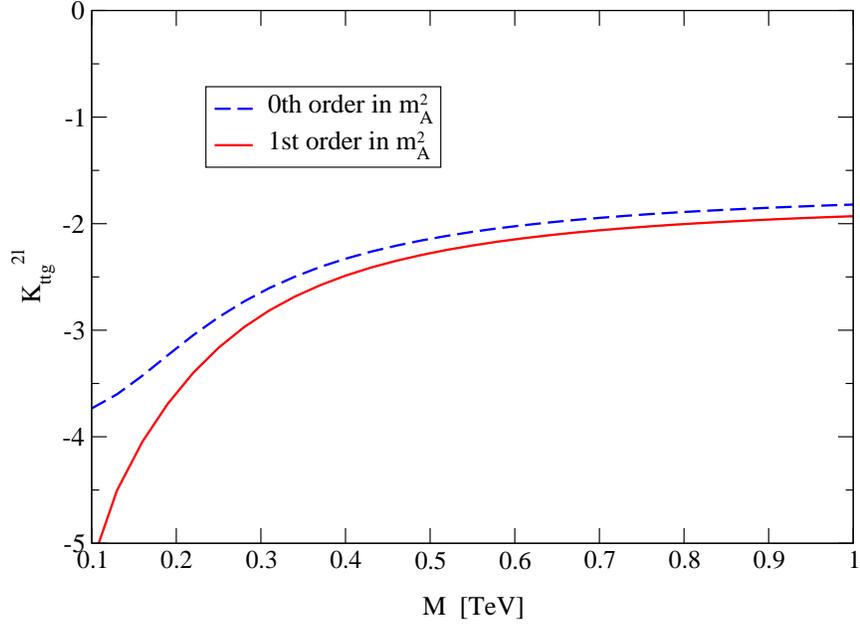}
\caption{Top-stop-gluino contribution $\HAt^{2\ell}$ as a function of
  a common SUSY mass $M$, for $\ma = 150$ GeV and $\tb=2$. The dashed
  line is the result in the limit of vanishing $\ma$, while the solid
  line includes the first-order term of a Taylor expansion in
  $\ma^2$.}
\label{fig:plot_top}
\end{center}
\end{figure}

\begin{figure}[p]
\begin{center}
\epsfig{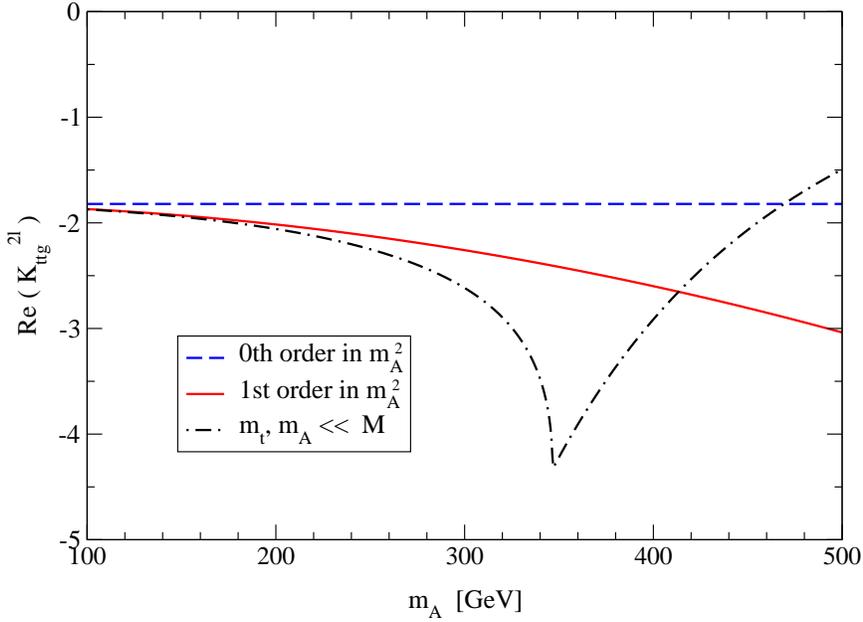}
\caption{Real part of $\HAt^{2\ell}$ as a function of $\ma$, for a
  common SUSY mass $M =1$ TeV and $\tb=2$. The solid and dashed lines
  are as in figure \ref{fig:plot_top} above, while the dot-dashed line is
  the result of an asymptotic expansion in $M$ which does not assume a
  specific hierarchy between $\mt$ and $\ma$.}
\label{fig:plot_top2}
\end{center}
\end{figure}

In figure \ref{fig:plot_top} we show the top-stop-gluino contribution
to the two-loop form factor for pseudoscalar production, i.e., the
term $\HAt^{2\ell}$ entering eq.~(\ref{HA2l}), as a function of the
common SUSY mass $M$, for $\ma = 150$ GeV and $\tb=2$. Even for the
lowest value of $M$ considered in the plot, $M=100$ GeV, the stop and
sbottom masses are above the threshold for real-particle
production. The dashed line represents the result obtained in the
limit of vanishing $\ma$, shown explicitly in eqs.~(\ref{compact}) and
(\ref{ffunc}), while the solid line represents the result computed at
the the first order of the Taylor expansion in the pseudoscalar mass,
i.e.~it includes the effect of terms of ${\cal O}(\ma^2/\mt^2)$ and
${\cal O}(\ma^2/M^2)$ which are too long to be presented in analytic
form. In the computation of these additional terms we assumed that the
${\cal O}(\ma^2/\mt^2)$ part of the one-loop top contribution, see
eq.~(\ref{K1lim1}), is expressed in terms of the pole top mass.

It can be seen in figure \ref{fig:plot_top} that the two-loop
top-stop-gluino contribution $\HAt^{2\ell}$ is of non-decoupling
nature, i.e., it does not tend to zero when all the superparticle
masses become large (note that the superpotential parameter $\mu$
increases together with the SUSY-breaking parameters). In addition,
the comparison between the solid and dashed lines shows that when the
common SUSY mass $M$ is close to $\ma$ the combined effect of the
terms of ${\cal O}(\ma^2/\mt^2)$ and ${\cal O}(\ma^2/M^2)$ can be as
large as 20\%--25\% with respect to the result obtained for vanishing
$\ma$.  However, when $M$ increases the effect of the terms of ${\cal
  O}(\ma^2/M^2)$ becomes quickly negligible. The remaining discrepancy
between the solid and dashed lines for moderate to large values of $M$
is due to the terms of ${\cal O}(\ma^2/\mt^2)$, and it amounts to a
modest 6\% for the value of $\ma$ considered in this example.

To assess the importance of the terms of ${\cal O}(\ma^2/\mt^2)$ for
larger values of $\ma$, we plot in figure \ref{fig:plot_top2} the real
part of $\HAt^{2\ell}$ as a function of the pseudoscalar mass, up to a
value $\ma$ = 500 GeV well above the threshold for real top-quark
production. The common SUSY mass is set to the relatively large value
$M=1$ TeV, and $\tb = 2$. As in figure \ref{fig:plot_top2}, the dashed
and solid lines represent the results obtained at the zeroth and first
order of the Taylor expansion in $\ma^2$, respectively.  The
comparison between those lines shows that when $\ma$ approaches
$2\,\mt$ the effect of the terms of ${\cal O}(\ma^2/\mt^2)$ gets as
large as 30\% with respect to the result obtained for vanishing
$\ma$. However, it is natural to wonder whether a Taylor expansion in
$\ma^2$ can give an accurate approximation to $\HAt^{2\ell}$ for
values of $\ma$ close to or larger than $\mt$. To address this
question, we show in figure \ref{fig:plot_top2} as a dot-dashed line
the result of the asymptotic expansion in $M$, given explicitly in
eqs.~(\ref{Kttg2})--(\ref{R4}). This result was derived under the
assumption that both $\ma$ and $\mt$ are much smaller than $M$, which
is indeed the case for $M=1$ TeV, but it does not require any specific
hierarchy between $\ma$ and $\mt$. The comparison between the
dot-dashed and solid lines shows that the Taylor expansion at the
first order in $\ma^2$ provides a good description of the dependence
of $\HAt^{2\ell}$ on the ratio $\ma/\mt$ up to values of $\ma$ of the
order of 250 GeV. On the other hand, when $\ma$ reaches the threshold
for real top production (i.e., at the cusp of the dot-dashed line) the
result of the asymptotic expansion in $M$ is roughly 80\% larger in
absolute value than the result at the first order of the Taylor
expansion in $\ma^2$, and a full 140\% larger than the result obtained
for vanishing $\ma$.

In summary, it appears that the compact result for $\HAt^{2\ell}$
given in eqs.~(\ref{compact}) and (\ref{ffunc}), which was derived for
$\ma=0$, can be safely applied only to scenarios in which $\ma$ is
smaller than $\mt$. While the inclusion of the terms proportional to
$\ma^2$ pushes the validity of the Taylor expansion up to larger
values of $\ma$, the expansion fails when $\ma$ gets close to the
threshold for real top production. In that case one can use the result
of the asymptotic expansion in $M$, provided that the latter is still
considerably larger than $\ma$.

We are now ready to discuss the relative importance of the various
two-loop contributions to the form factor for pseudoscalar production.
We will see that, at least in the region of the parameter space that
we consider in this example, the results are qualitatively similar to
what we found in ref.~\cite{noibot} for the case of the heavy scalar
$H$.

A precise NLO determination of the cross section for pseudoscalar
production would require us to take into account the contribution of
one-loop diagrams with real parton emission, and to perform an
integration over the phase space (see section
\ref{sec:general}). However, for the purpose of illustrating the
relative importance of the various two-loop contributions, we can just
define a factor $K_\smalla$ that contains the ratio of two-loop to
one-loop form factors appearing in eq.~(\ref{ggg}):
\be
\label{kfac}
K_\smalla ~=~ 1\,+ \,2\,\frac{\alpha_s}{\pi}\,{\rm Re}
\left(\frac{\HA^{2\ell}}{\HA^{1\ell}}\right)~.
\ee

\begin{figure}[t]
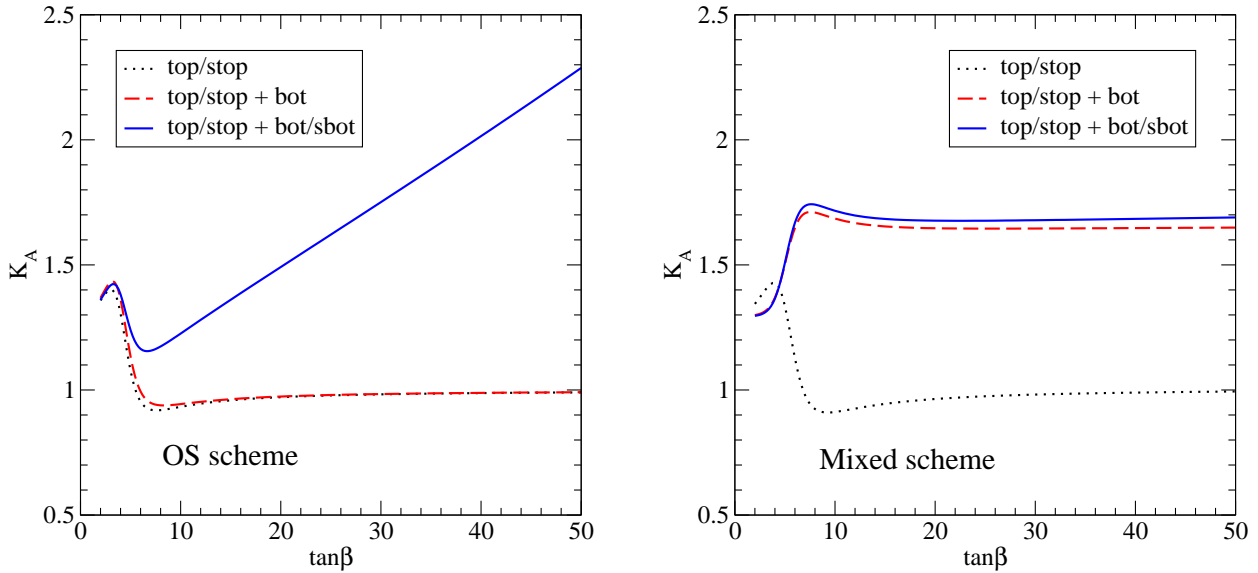

\begin{center}
\mbox{
\epsfig{figure=KfacOSA.eps,width=7.8cm}~~~~~~
\epsfig{figure=KfacORA.eps,width=7.8cm}
}
\caption{$K$ factor for the production of a pseudoscalar Higgs $A$ as
  a function of $\tan\beta$, for $m_A = 150$ GeV and all SUSY mass
  parameters equal to $M=500$ GeV. The three lines show the effect of
  the different two-loop contributions, in the OS scheme (left panel)
  and in the ``mixed'' scheme (right panel).}
\label{fig:kfactor}
\end{center}
\end{figure}

In the left panel of figure \ref{fig:kfactor} we plot $K_\smalla$ as a
function of $\tan\beta$, for $m_A = 150$ GeV and all SUSY mass
parameters equal to $M=500$ GeV. The one-loop form factor
$\HA^{1\ell}$ in eq.~(\ref{kfac}) contains both the top and bottom
contributions, computed under the approximations of
eqs.~(\ref{K1lim1}) and (\ref{K1lim2}), respectively. We identify the
quark masses in the one-loop form factor with the pole masses, and
refer to this choice as ``on-shell'' (OS) scheme.  The lines in the
plot correspond to different computations of the two-loop form factor
$\HA^{2\ell}$: the dotted line includes only the contributions of the
top/stop sector (both those involving top quarks and gluons and those
involving top, stop and gluinos) computed at the first order of the
Taylor expansion in $\ma^2$; the dashed line includes also the
contributions of two-loop diagrams with bottom quarks and gluons;
finally, the solid line includes the full contributions of the
bottom/sbottom sector.

Comparing the three lines in the left panel of figure
\ref{fig:kfactor} it can be seen that the top/stop contributions
dominate the two-loop form factor up to values of $\tb$ around 5.  For
larger values of $\tan\beta$ the contribution of the
bottom-sbottom-gluino diagrams (included in the solid line) becomes
the dominant one, and $K_\smalla$ grows linearly with $\tb$.  This
behavior can be understood by recalling that, as can be seen in
eq.~(\ref{couplings}), the Yukawa coupling of the pseudoscalar to
bottom quarks is enhanced by $\tan\beta$ with respect to the coupling
of the SM Higgs, while the coupling to top quarks is suppressed by
$\tan\beta$.  Consequently, for moderate to large values of $\tb$ both
the one-loop and the two-loop form factors in $K_\smalla$ are
dominated by the contribution of the diagrams controlled by the
pseudoscalar-bottom coupling, with the result that the coupling itself
cancels out in the ratio. However, the dominant contribution from the
bottom-sbottom-gluino diagrams in the OS scheme, see
eq.~(\ref{bsbg2}), contains an additional $\tan\beta$-enhancement,
which explains the linear rise of $K_\smalla$. On the other hand, the
proximity between the dotted and dashed lines shows that, in the OS
scheme, the contribution to $\HA^{2\ell}$ of the two-loop diagrams
with bottom quarks and gluons is very small. This is due to a partial
cancellation among the three terms entering ${\cal K}_{bg}^{2\ell}$ in
eq.~(\ref{K2lonshell}), and to the fact that, in this scheme, the term
${\cal F}_2(\tau_b)$ is not enhanced by the potentially large
logarithm of the ratio between the bottom mass and the renormalization
scale, as can be seen by comparing eqs.~(\ref{K2lonshell}) and
(\ref{K2ldrbar}).

As discussed in section \ref{sec:botsbot}, all $\tb$-enhanced terms
cancel out in a ``mixed'' renormalization scheme in which the
pseudoscalar-bottom Yukawa coupling in the one-loop part of the result
is identified with the $\drbar$-renormalized MSSM bottom mass
$\widehat m_b(Q)$, where $Q$ is a reference scale that we take equal
to $\ma$, while the mass of the bottom quark running in the loop is
identified with the pole mass $M_b$. To determine $\widehat m_b(\ma)$,
we first evolve the $\msbar$-renormalized SM mass $\overline \mb(\mb)$
up to the scale $\ma$ via the NLO-QCD renormalization group equations,
then we convert it to the $\drbar$-renormalized SM mass $\widehat
m_b^{\smallsm}(\ma)$ via the appropriate shift, and finally we convert
it to the MSSM running mass according to
\be
\label{mbrun}
\widehat \mb(\ma) ~=~ \widehat \mb^{\smallsm}(\ma)\,
\frac{1+\delta_b}{~1+\Delta_b}~, 
\ee
where $\Delta_b$ is given in eq.~(\ref{deltab}), and $\delta_b$ is
proportional to the part of $(\delta m_b)^{\scriptscriptstyle SUSY}$
in eq.~(\ref{dmbovmb}) that is not enhanced by $\tan\beta$:
\be
\label{db}
\delta_b ~=~ -\frac{\as\,C_F}{4\pi}\,\left[
\ln\frac{\mg^2}{\ma^2}  + f(x_1)+f(x_2) + \frac{2\,\mg\,A_b}{\bu-\bd}\,
\left(\frac{x_1}{1-x_1} \ln x_1-\frac{x_2}{1-x_2}\ln x_2\right)\right]~.
\ee

The ``mixed'' renormalization prescription is realized by computing
the one-loop bottom contribution ${\cal K}^{1\ell}(\tau_b)$ in
eq.~(\ref{H1lp}) in terms of the pole mass $M_b$, then rescaling it by
a factor $\widehat m_b(\ma)/M_b$. The two-loop form factor
$\HA^{2\ell}$ must then be shifted as in eq.~(\ref{shift}).  In the
right panel of fig.~\ref{fig:kfactor} we present the result of this
manipulation.  The input parameters and the meaning of the different
lines are the same as for the plot in the left panel.  The proximity
between the dashed and solid lines, and the flatness of the lines for
moderate to large values of $\tb$, show that the contribution of the
two-loop bottom-sbottom-gluino diagrams is rather small in this
renormalization scheme, and it does not induce an additional
$\tan\beta$-enhancement. However, the comparison between the dotted
and dashed lines shows that there is a sizable contribution to
$K_\smalla$ from the two-loop diagrams involving bottom quarks and
gluons. This is due to the fact that the shift in eq.~(\ref{shift})
brings back a large logarithm, $\ln (\mb^2/\ma^2)$, which compensates
the scale dependence of the running mass $\widehat m_b$.

%%%%%%%%%%%%%%%%%%%%%%%%%%%%%%%%%%%%%%%%%%%%%%%%%%%%%%%%%%%%%

\section{Conclusions}
\label{sec:concl}

The calculation of the production cross section for the MSSM Higgs
bosons is not quite as advanced as in the SM. Indeed, despite valiant
efforts \cite{babis2,spiraDb}, a full computation of the two-loop
quark-squark-gluino contributions, valid for arbitrary values of all
the relevant particle masses, has not been made publicly available so
far. Approximate analytic results, however, can be derived if the
Higgs bosons are somewhat lighter than the squarks and the gluinos. In
the MSSM this condition almost certainly applies to the lightest
scalar $h$. Moreover, recent results from SUSY searches at the LHC
\cite{atlas} set preliminary lower bounds on the squark and gluino
masses just below the TeV (albeit for specific models of SUSY
breaking), suggesting that there might be wide regions of the MSSM
parameter space in which the condition also applies to the heavy
scalar $H$ and to the pseudoscalar $A$.

In this paper we presented a calculation of the two-loop
quark-squark-gluino contributions to the cross section for
pseudoscalar production. We exploited techniques developed in our
earlier computations of the production cross section for the CP-even
Higgs bosons of the MSSM~\cite{DS,noibot} to obtain explicit and
compact analytic results based on expansions in the heavy particle
masses. We avoided problems related to the definition of the Dirac
matrix $\gamma_5$ in $n_d \ne 4$ dimensions, which are specific to the
case of pseudoscalar production, by regularizing the loop integrals
with the Pauli-Villars method. For what concerns the top-stop-gluino
contributions, we provided both the result of a Taylor expansion in
the pseudoscalar mass, up to and including terms of ${\cal
  O}(\ma^2/\mt^2)$ and ${\cal O}(\ma^2/M^2)$, and the result of an
asymptotic expansion in the superparticle masses, up to and including
terms of ${\cal O}(\ma^2/M^2)$ and ${\cal O}(\mt^2/M^2)$.  The latter
can be easily adapted to the case of the bottom-sbottom-gluino
contributions, providing a result valid up to and including terms of
${\cal O}(\mb^2/\ma^2)$ and ${\cal O}(\mb/M)$. We discussed how the
$\tb$-enhanced terms in the bottom-sbottom-gluino contributions can be
eliminated via an appropriate choice of renormalization scheme for the
parameters entering the one-loop part of the calculation, and compared
our results with those obtained in the effective-Lagrangian
approximation. All of our results can be easily implemented in
computer codes for an efficient and accurate determination of the
cross section for pseudoscalar production.

Finally, the results derived in this paper for the production cross
section can be straightforwardly adapted to the NLO computation of the
gluonic and photonic decay widths of the pseudoscalar Higgs boson in
the MSSM, in analogy to what described in section 5 of ref.~\cite{DS}
for the case of the CP-even bosons.

%%%%%%%%%%%%%%%%%%%%%%%%%%%%%%%%%%%%%%%%%%%%%%%%%%%%%%%%%%%%%

\section*{Acknowledgments}
We thank M.~Spira for useful communications about ref.~\cite{SDGZ}.
This work was partly supported by the Research Executive Agency (REA)
of the European Union through the Initial Training Network LHCPhenoNet
under contract PITN-GA-2010-264564.  The diagrams in figure
\ref{fig:diags} were drawn using JaxoDraw \cite{bt}.

%%%%%%%%%%%%%%%%%%%%%%%%%%%%%%%%%%%%%%%%%%%%%%%%%%%%%%%%%%%%%%
\vfill
\newpage

\section*{Appendix: NLO contributions from real parton emission}
\label{appA}
\begin{appendletterA}

In this appendix we present for completeness our results for the NLO
contributions to pseudoscalar production from one-loop diagrams with
emission of a real parton, i.e., the functions ${\cal R}_{gg},\, {\cal
  R}_{q \bar q}$ and ${\cal R}_{q g}$ entering eqs.~(\ref{ggg}) and
(\ref{qqqg}). Such contributions were first computed in
ref.~\cite{SDGZ} (see also ref.~\cite{FDS}).

The contribution of the gluon-fusion channel, $g g \to A g $, can be
written as
\be
{\cal R}_{gg} = \frac1{z(1-z)}\int_0^1 \frac{d v}{v (1-v)} \left\{
 8\,z^4\, \frac{\left| {\cal A}_{gg}(\hat{s},\hat{t},\hat{u})\right|^2}{\left| 
   \HA^{1\ell} \right|^2 }  - 
(1-z+z^2)^2 \right\}~,
\ee
where $\hat{t} = -\hat{s}\, (1-z) (1-v),\,\hat{u} = -\hat{s}
\,(1-z)\,v$, and
\be
\left| {\cal A}_{gg}(s,t,u)\right|^2 ~=~ T_F^2 \, \left[ \cot^2 \beta
\left| {\cal A}^{tt}_{gg}(s,t,u)\right|^2  +
\tan^2 \beta \left| {\cal A}^{bb}_{gg}(s,t,u)\right|^2  +
2 \left| {\cal A}^{tb}_{gg}(s,t,u)\right|^2  \right]~,
\ee
with 
\be
\left| {\cal A}^{ij}_{gg}(s,t,u) \right|^2  =
 |A^{ij} (s,t,u)|^2 + |A^{ij} (u,s,t)|^2 + |A^{ij} (t,u,s)|^2 ~.
\label{Agg}
\ee
Defining, for $i=t,b$~,
\be
y_i \equiv \frac{m_i^2}{\ma^2}, ~~~~~~
s_i \equiv \frac{s}{m_i^2},~~~~~t_i \equiv \frac{t}{m_i^2}, ~~~~~~~
u_i \equiv \frac{u}{m_i^2},
\label{defx}
\ee
we find:
\bea
|A^{ij} (s,t,u)|^2 & =  & \frac{y_i y_j}{4\ma^4} \,\biggr\{\biggr[~~~
b_1(s,t,u)\, H_2(s_i,y_i)\,H_2^\dagger(s_j,y_j) +
b_2(s,t,u)\, H_2(s_i,y_i)\,H_2^\dagger(t_j,y_j) \nn \\
& &  ~~~~~~~~~ +     
b_3(s,t,u)\,H_3(s_i,t_i,u_i)\,H_3^\dagger(s_j,t_j,u_j) +
b_4(s,t,u)\,H_3(s_i,t_i,u_i)\,H_3^\dagger(u_j,s_j,t_j)\nn \\
& &  ~~~~~~~~~ +     
b_5(s,t,u)\,H_2(s_i,y_i)\,H_3^\dagger(s_j,t_j,u_j) +
b_6(s,t,u)\,H_2(s_i,y_i)\,H_3^\dagger(t_j,u_j,s_j)\nn \\
& &  ~~~~~~~~~ +
b_7(s,t,u)\,H_2(s_i,y_i)\,H_3^\dagger(u_j,s_j,t_j)\biggr] ~+~ 
(i \leftrightarrow j) ~\biggr\} ~+~ {\rm h. c.}~,
\label{A24fun}
\eea
where the function $H_3(s,t,u)$ is defined in eq.~(2.28) of
ref.~\cite{BDV}, and
\be
H_2(s,y) = \frac{1}{2} \left[ \log^2 
\left( \frac{\sqrt{1- 4/s} - 1}{\sqrt{1- 4/s} + 1} \right) -
\log^2  \left( \frac{\sqrt{1- 4 y} - 1}{\sqrt{1- 4y } + 1} \right)
\right]~.
\ee
The coefficient functions $b_i(s,t,u)$ entering eq.~(\ref{A24fun}) are
\bea
b_{1}(s,t,u) &=& \frac1{2}\, \biggr[ \frac{4 t^2 u^2}{(t+u)^2} 
+  s^2 - 3 t u + s (t + u) + (t+u)^2 \biggr]~, \\
b_{2}(s,t,u) &=&  s^2 + t^2 + u^2 +  s t + 
            \frac{2 s^2 t u}{(s-t)(s+u)} - \frac{2 s t^2 u}{(s-t)(t+u)}
                ~, \\
b_{3}(s,t,u) &=& \frac1{8}\, \biggr[ s^2 + t^2 + u^2 +  t u +
                  s (t + u)\, \biggr]~, \\ 
b_{4}(s,t,u) &=&  \frac1{4}\, (s+t) (t + u)~, \\
b_{5}(s,t,u) &=& -\frac1{2}\, \biggr[t^2 + u^2 + s (t + u)\biggr]~, \\ 
b_{6}(s,t,u) &=& -\frac1{2}\, \biggr[s^2 + (t +u)(s+u)+
                      \frac{(t-u) u t}{(t + u)} \biggr] ~, \\ 
b_{7}(s,t,u) &=& -\frac1{2}\, \biggr[ s^2 + (t +u)(s+t)+
                      \frac{(u-t) u t}{(t + u)} \biggr]~.
\eea 
 
The contribution of the quark-antiquark annihilation channel, $ q \bar
q \to A g $, can be written as
\be
{\cal R}_{q \bar q} = \frac{512}{27} 
 \frac{z\,(1-z)\, \left| {\cal A}_{q \bar q}(\hat{s})\right|^2}
{\left|\HA^{1\ell}\right|^2 }  \, ,
\ee
with
\be
{\cal A}_{q \bar q} (s)  =   
       T_F \, \left[ \cot \beta \, y_t  \, H_2^{\phantom{\dagger}} (s_t,y_t) +
                     \tan \beta \, y_b  \, H_2(s_b,y_b) \right]~.  
\ee

Finally, the contribution of the  quark-gluon scattering
channel, $ q g \to A q $, can be written as
\be
{\cal R}_{qg} ~=~ \frac{C_F}2 z~+~ C_F  \int_0^1  \frac{d v}{(1-v)} \left\{
\frac{ 1 + (1-z)^2 v^2}{[1-(1-z) v]^2} \,
\frac{8 \,z \left| {\cal A}_{q\bar q}(\hat{t}\,)\right|^2 }{  
\left|\HA^{1\ell} \right|^2 } ~-~ \frac{1+(1 \! - \! z)^2}{2 z} \right\}~.
\ee

We compared our results for the functions ${\cal R}_{gg},\, {\cal
  R}_{q \bar q}$ and ${\cal R}_{q g}$ with the corresponding results
in ref.~\cite{SDGZ}, and found full agreement.\footnote{Some misprints
  in ref.~\cite{SDGZ} must be taken into account in the comparison. In
  eq.~(C.4) of that paper the term within square modulus in the
  definition of $d_{gq}$ should be divided by 2. Also, the formulae in
  the Appendices B and C omit all occurrences of the MSSM Higgs-quark
  couplings, denoted in that paper as $g_Q^\Phi$.}

\end{appendletterA}

%%%%%%%%%%%%%%%%%%%%%%%%%%%%%%%%%%%%%%%%%%%%%%%%%%%%%%%%%%%%%%


\vfill
\newpage

\begin{thebibliography}{99}

\bibitem{H2gQCD0} H.~M.~Georgi, S.~L.~Glashow, M.~E.~Machacek and
  D.~V.~Nanopoulos,
  %``Higgs Bosons From Two Gluon Annihilation In Proton Proton Collisions,''
  Phys.\ Rev.\ Lett.\  {\bf 40} (1978) 692.
  %%CITATION = PRLTA,40,692;%%

\bibitem{H2gQCD1}
  S.~Dawson,
  %``Radiative Corrections To Higgs Boson Production,''
  Nucl.\ Phys.\ B {\bf 359} (1991) 283;
  %%CITATION = NUPHA,B359,283;%%
%
  A.~Djouadi, M.~Spira and P.~M.~Zerwas,
  %``Production of Higgs bosons in proton colliders: QCD corrections,''
  Phys.\ Lett.\ B {\bf 264} (1991) 440.
  %%CITATION = PHLTA,B264,440;%%

\bibitem{SDGZ}
  M.~Spira, A.~Djouadi, D.~Graudenz and P.~M.~Zerwas,
  %``Higgs boson production at the LHC,''
  Nucl.\ Phys.\ B {\bf 453} (1995) 17
  [arXiv:hep-ph/9504378].
  %%CITATION = HEP-PH 9504378;%%

\bibitem{HK0}
  R.~Harlander and P.~Kant,
  %``Higgs production and decay: Analytic results at next-to-leading order
  %QCD,''
  JHEP {\bf 0512} (2005) 015
  [arXiv:hep-ph/0509189].
  %%CITATION = HEP-PH 0509189;%%

 \bibitem{H2gQCD2}
  R.~V.~Harlander,
  %``Virtual corrections to g g $\to$ H to two loops in the heavy top limit,''
  Phys.\ Lett.\ B {\bf 492} (2000) 74.
  [arXiv:hep-ph/0007289];
  %%CITATION = HEP-PH 0007289;%%
%
  S.~Catani, D.~de Florian and M.~Grazzini,
 %``Higgs production in hadron collisions: Soft and virtual QCD corrections  at
  %NNLO,''
  JHEP {\bf 0105} (2001) 025
  [arXiv:hep-ph/0102227];
  %%CITATION = HEP-PH 0102227;%%
%
  R.~V.~Harlander and W.~B.~Kilgore,
  %``Soft and virtual corrections to p p $\to$ H + X at NNLO,''
  Phys.\ Rev.\ D {\bf 64} (2001) 013015
  [arXiv:hep-ph/0102241],
  %%CITATION = HEP-PH 0102241;%%
%
  %``Next-to-next-to-leading order Higgs production at hadron colliders,''
  Phys.\ Rev.\ Lett.\  {\bf 88} (2002) 201801
  [arXiv:hep-ph/0201206];
  %%CITATION = HEP-PH 0201206;%%
%
  C.~Anastasiou and K.~Melnikov,
  %``Higgs boson production at hadron colliders in NNLO QCD,''
  Nucl.\ Phys.\ B {\bf 646} (2002) 220
  [arXiv:hep-ph/0207004];
  %%CITATION = HEP-PH 0207004;%%
%
  V.~Ravindran, J.~Smith and W.~L.~van Neerven,
  %``NNLO corrections to the total cross section for Higgs boson production  in
  %hadron hadron collisions,''
  Nucl.\ Phys.\ B {\bf 665} (2003) 325
  [arXiv:hep-ph/0302135].
  %%CITATION = HEP-PH 0302135;%%

\bibitem{H2gQCD3}
S.~Marzani, R.~D.~Ball, V.~Del Duca, S.~Forte and A.~Vicini,
  %``Higgs production via gluon-gluon fusion with finite top mass beyond
  %next-to-leading order,''
  Nucl.\ Phys.\  B {\bf 800} (2008) 127
  [arXiv:0801.2544 [hep-ph]];
  %%CITATION = NUPHA,B800,127;%%
%
  S.~Marzani, R.~D.~Ball, V.~Del Duca, S.~Forte and A.~Vicini,
  %``Finite-top-mass effects in NNLO Higgs production,''
  Nucl.\ Phys.\ Proc.\ Suppl.\  {\bf 186} (2009) 98
  [arXiv:0809.4934 [hep-ph]];
  %%CITATION = NUPHZ,186,98;%%
%
R.~V.~Harlander and K.~J.~Ozeren,
  %``Top mass effects in Higgs production at next-to-next-to-leading order QCD:
  %virtual corrections,''
  Phys.\ Lett.\  B {\bf 679} (2009) 467
  [arXiv:0907.2997 [hep-ph]], 
  %%CITATION = PHLTA,B679,467;%%
  JHEP {\bf 0911} (2009) 088
  [arXiv:0909.3420 [hep-ph]];
  %%CITATION = JHEPA,0911,088;%%
%
 A.~Pak, M.~Rogal and M.~Steinhauser,
  %``Virtual three-loop corrections to Higgs boson production in gluon fusion
  %for finite top quark mass,''
  Phys.\ Lett.\  B {\bf 679} (2009) 473
  [arXiv:0907.2998 [hep-ph]],
  %%CITATION = PHLTA,B679,473;%%
  JHEP {\bf 1002} (2010) 025
  [arXiv:0911.4662 [hep-ph]];
  %%CITATION = JHEPA,1002,025;%%
%
 R.~V.~Harlander, H.~Mantler, S.~Marzani and K.~J.~Ozeren,
  %``Higgs production in gluon fusion at next-to-next-to-leading order QCD for
  %finite top mass,''
  Eur.\ Phys.\ J.\  C {\bf 66} (2010) 359
  [arXiv:0912.2104 [hep-ph]].
  %%CITATION = EPHJA,C66,359;%%

\bibitem{bd4}
  S.~Catani, D.~de Florian, M.~Grazzini and P.~Nason,
  %``Soft-gluon resummation for Higgs boson production at hadron colliders,''
  JHEP {\bf 0307} (2003) 028
  [arXiv:hep-ph/0306211].
  %%CITATION = HEP-PH 0306211;%%

\bibitem{Moch:2005ky}
  S.~Moch and A.~Vogt,
  %``Higher-order soft corrections to lepton pair and Higgs boson production,''
  Phys.\ Lett.\  B {\bf 631} (2005) 48
  [arXiv:hep-ph/0508265];
  %%CITATION = PHLTA,B631,48;%%
%
  V.~Ravindran,
  %``Higher-order threshold effects to inclusive processes in QCD,''
  Nucl.\ Phys.\  B {\bf 752} (2006) 173
  [arXiv:hep-ph/0603041].
  %%CITATION = NUPHA,B752,173;%%

\bibitem{DjG}
  A.~Djouadi and P.~Gambino,
  %``Leading electroweak correction to Higgs boson production at proton
  %colliders,''
  Phys.\ Rev.\ Lett.\  {\bf 73} (1994) 2528
  [arXiv:hep-ph/9406432];
  %%CITATION = HEP-PH 9406432;%%
% 
 A.~Djouadi, P.~Gambino and B.~A.~Kniehl,
  %``Two-loop electroweak heavy-fermion corrections to Higgs-boson  production
  %and decay,''
  Nucl.\ Phys.\ B {\bf 523} (1998) 17
  [arXiv:hep-ph/9712330].
  %%CITATION = HEP-PH 9712330;%%

\bibitem{ABDV0}
  U.~Aglietti, R.~Bonciani, G.~Degrassi and A.~Vicini,
  %``Two-loop light fermion contribution to Higgs production and decays,''
  Phys.\ Lett.\ B {\bf 595} (2004) 432
  [arXiv:hep-ph/0404071],
  %%CITATION = HEP-PH 0404071;%%
%
  %  U.~Aglietti, R.~Bonciani, G.~Degrassi and A.~Vicini,
 %``Master integrals for the two-loop light fermion contributions to g g  $\to$
  %H and H $\to$ gamma gamma,''
  Phys.\ Lett.\ B {\bf 600} (2004) 57
  [arXiv:hep-ph/0407162];
  %%CITATION = HEP-PH 0407162;%%
%
  G.~Degrassi and F.~Maltoni,
  %``Two-loop electroweak corrections to Higgs production at hadron
  %colliders,''
  Phys.\ Lett.\ B {\bf 600} (2004) 255
  [arXiv:hep-ph/0407249].
  %%CITATION = HEP-PH 0407249;%%

\bibitem{APSU}
 S.~Actis, G.~Passarino, C.~Sturm and S.~Uccirati,
  %``NLO Electroweak Corrections to Higgs Boson Production at Hadron
  %Colliders,''
  Phys.\ Lett.\  B {\bf 670} (2008) 12
  [arXiv:0809.1301 [hep-ph]],
  %%CITATION = PHLTA,B670,12;%%
% S.~Actis, G.~Passarino, C.~Sturm and S.~Uccirati,
  %``NNLO Computational Techniques: the Cases $H \to \gamma \gamma$ and $H \to g
  %g$,''
  Nucl.\ Phys.\  B {\bf 811} (2009) 182
  [arXiv:0809.3667 [hep-ph]].
  %%CITATION = NUPHA,B811,182;%%

\bibitem{NNLOSUSY}
  A.~Pak, M.~Steinhauser and N.~Zerf,
  %``Towards Higgs boson production in gluon fusion to NNLO in the MSSM,''
  Eur.\ Phys.\ J.\  C {\bf 71} (2011) 1602
  [arXiv:1012.0639 [hep-ph]].
  %%CITATION = EPHJA,C71,1602;%%

\bibitem{Dawson:1996xz}
  S.~Dawson, A.~Djouadi and M.~Spira,
  %``QCD Corrections to SUSY Higgs Production: The Role of Squark Loops,''
  Phys.\ Rev.\ Lett.\  {\bf 77} (1996) 16
  [arXiv:hep-ph/9603423].
  %%CITATION = PRLTA,77,16;%%

\bibitem{HS}
  R.~V.~Harlander and M.~Steinhauser,
  %``Hadronic Higgs production and decay in supersymmetry at next-to-leading
  %order,''
  Phys.\ Lett.\  B {\bf 574} (2003) 258
  [arXiv:hep-ph/0307346],
  %%CITATION = PHLTA,B574,258;%%
%
  %R.~V.~Harlander and M.~Steinhauser,
  %``Supersymmetric Higgs production in gluon fusion at next-to-leading
  %order,''
  JHEP {\bf 0409} (2004) 066
  [arXiv:hep-ph/0409010].
  %%CITATION = JHEPA,0409,066;%%

\bibitem{DS}
  G.~Degrassi and P.~Slavich,
  %``On the NLO QCD corrections to Higgs production and decay in the MSSM,''
  Nucl.\ Phys.\  B {\bf 805} (2008) 267
  [arXiv:0806.1495 [hep-ph]].
  %%CITATION = NUPHA,B805,267;%%

\bibitem{babis1}
  C.~Anastasiou, S.~Beerli, S.~Bucherer, A.~Daleo and Z.~Kunszt,
  %``Two-loop amplitudes and master integrals for the production of a Higgs
  %boson via a massive quark and a scalar-quark loop,''
  JHEP {\bf 0701} (2007) 082
  [arXiv:hep-ph/0611236].
  %%CITATION = JHEPA,0701,082;%%

\bibitem{ABDV}
  U.~Aglietti, R.~Bonciani, G.~Degrassi and A.~Vicini,
  %``Analytic results for virtual QCD corrections to Higgs production and
  %decay,''
  JHEP {\bf 0701} (2007) 021
  [arXiv:hep-ph/0611266].
  %%CITATION = JHEPA,0701,021;%%

\bibitem{MS}
  M.~Muhlleitner and M.~Spira,
  %``Higgs boson production via gluon fusion: Squark loops at NLO QCD,''
  Nucl.\ Phys.\  B {\bf 790} (2008) 1
  [arXiv:hep-ph/0612254].
  %%CITATION = NUPHA,B790,1;%%

\bibitem{BDV} R.~Bonciani, G.~Degrassi and A.~Vicini,
  %``Scalar Particle Contribution to Higgs Production via Gluon Fusion 
  % at NLO,''
  JHEP {\bf 0711} (2007) 095
  [arXiv:0709.4227 [hep-ph]].
  %%CITATION = JHEPA,0711,095;%%

\bibitem{KLS}
  M.~Kramer, E.~Laenen and M.~Spira,
  %``Soft gluon radiation in Higgs boson production at the LHC,''
  Nucl.\ Phys.\ B {\bf 511}, 523 (1998)
  [arXiv:hep-ph/9611272].
  %%CITATION = HEP-PH 9611272;%%

\bibitem{babis2}
  C.~Anastasiou, S.~Beerli and A.~Daleo,
  %``The two-loop QCD amplitude gg -> h,H in the Minimal Supersymmetric Standard
  %Model,''
  Phys.\ Rev.\ Lett.\  {\bf 100} (2008) 241806
  [arXiv:0803.3065 [hep-ph]].
  %%CITATION = PRLTA,100,241806;%%

\bibitem{spiraDb}
  M.~Muhlleitner, H.~Rzehak and M.~Spira,
  %``SUSY-QCD corrections to MSSM Higgs boson production via gluon fusion,''
  arXiv:1001.3214 [hep-ph].
  %%CITATION = ARXIV:1001.3214;%%

\bibitem{noibot}
  G.~Degrassi and P.~Slavich,
  %``NLO QCD bottom corrections to Higgs boson production in the MSSM,''
  JHEP {\bf 1011} (2010) 044
  [arXiv:1007.3465 [hep-ph]].
  %%CITATION = JHEPA,1011,044;%%

\bibitem{hhm}
  R.~V.~Harlander, F.~Hofmann and H.~Mantler,
  %``Supersymmetric Higgs production in gluon fusion,''
  JHEP {\bf 1102} (2011) 055
  [arXiv:1012.3361 [hep-ph]].
  %%CITATION = JHEPA,1102,055;%%

\bibitem{AQCD}
  R.~P.~Kauffman and W.~Schaffer,
  %``QCD corrections to production of Higgs pseudoscalars,''
  Phys.\ Rev.\  D {\bf 49} (1994) 551
  [arXiv:hep-ph/9305279];
  %%CITATION = PHRVA,D49,551;%%
%
   M.~Spira, A.~Djouadi, D.~Graudenz and P.~M.~Zerwas,
  %``SUSY Higgs production at proton colliders,''
  Phys.\ Lett.\  B {\bf 318} (1993) 347.
  %%CITATION = PHLTA,B318,347;%%

\bibitem{CKSB}
  K.~G.~Chetyrkin, B.~A.~Kniehl, M.~Steinhauser and W.~A.~Bardeen,
  %``Effective QCD interactions of CP odd Higgs bosons at three loops,''
  Nucl.\ Phys.\  B {\bf 535} (1998) 3
  [arXiv:hep-ph/9807241].
  %%CITATION = NUPHA,B535,3;%%

\bibitem{CM}
  F.~Caola and S.~Marzani,
  %``Finite fermion mass effects in pseudoscalar Higgs production via
  %gluon-gluon fusion,''
  Phys.\ Lett.\  B {\bf 698} (2011) 275
  [arXiv:1101.3975 [hep-ph]].
  %%CITATION = PHLTA,B698,275;%%

\bibitem{HH}
  R.~V.~Harlander and F.~Hofmann,
  %``Pseudo-scalar Higgs production at next-to-leading order SUSY-QCD,''
  JHEP {\bf 0603} (2006) 050
  [arXiv:hep-ph/0507041].
  %%CITATION = JHEPA,0603,050;%%

\bibitem{effL}
  M.~S.~Carena, D.~Garcia, U.~Nierste and C.~E.~M.~Wagner,
  %``Effective Lagrangian for the $\bar{t} b H^{+}$ interaction in the MSSM and
  %charged Higgs phenomenology,''
  Nucl.\ Phys.\  B {\bf 577} (2000) 88
  [arXiv:hep-ph/9912516].
  %%CITATION = NUPHA,B577,88;%%

\bibitem{GHS}
  J.~Guasch, P.~Hafliger and M.~Spira,
  %``MSSM Higgs decays to bottom quark pairs revisited,''
  Phys.\ Rev.\  D {\bf 68} (2003) 115001
  [arXiv:hep-ph/0305101].
  %%CITATION = PHRVA,D68,115001;%%

\bibitem{HV}
  G.~'t Hooft and M.~J.~G.~Veltman,
  %``Regularization and Renormalization of Gauge Fields,''
  Nucl.\ Phys.\  B {\bf 44} (1972) 189.
  %%CITATION = NUPHA,B44,189;%%

\bibitem{larin}
  S.~A.~Larin,
  %``The Renormalization of the axial anomaly in dimensional regularization,''
  Phys.\ Lett.\  {\bf B303 } (1993)  113-118.
  [hep-ph/9302240].

\bibitem{BFM}
  B.~S.~Dewitt,
  %``Quantum Theory Of Gravity. Ii. The Manifestly Covariant Theory,''
  Phys.\ Rev.\  {\bf 162} (1967) 1195 ;
  %%CITATION = PHRVA,162,1195;%%
 %
  J.~Honerkamp,
  %``The Question Of Invariant Renormalizability Of The Massless
  %Yang-Mills Theory In A Manifest Covariant Approach,''
  Nucl.\ Phys.\ B {\bf 48} (1972) 269 ;
  %%CITATION = NUPHA,B48,269;%%
%
  H.~Kluberg-Stern and J.~B.~Zuber,
 %``Renormalization Of Nonabelian Gauge Theories In A Background Field Gauge: 1.
  %Green Functions,''
  Phys.\ Rev.\ D {\bf 12} (1975) 482;
  %%CITATION = PHRVA,D12,482;%%
%
  L.~F.~Abbott,
  %``The Background Field Method Beyond One Loop,''
  Nucl.\ Phys.\ B {\bf 185} (1981) 189.
  %%CITATION = NUPHA,B185,189;%%

\bibitem{dedeslav}
  A.~Dedes and P.~Slavich,
  %``Two loop corrections to radiative electroweak symmetry breaking in the
  %MSSM,''
  Nucl.\ Phys.\  B {\bf 657} (2003) 333
  [arXiv:hep-ph/0212132].
  %%CITATION = NUPHA,B657,333;%%

\bibitem{adler}
  S.~L.~Adler and W.~A.~Bardeen,
  %``Absence of higher order corrections in the anomalous axial vector
  %divergence equation,''
  Phys.\ Rev.\  {\bf 182} (1969) 1517.
  %%CITATION = PHRVA,182,1517;%%

\bibitem{hrs}
  L.~J.~Hall, R.~Rattazzi and U.~Sarid,
  %``The Top quark mass in supersymmetric SO(10) unification,''
  Phys.\ Rev.\  D {\bf 50} (1994) 7048
  [arXiv:hep-ph/9306309].
  %%CITATION = PHRVA,D50,7048;%%

\bibitem{PDG}
  K.~Nakamura {\it et al.}  [Particle Data Group],
  %``Review of particle physics,''
  J.\ Phys.\ G {\bf 37} (2010) 075021.
  %%CITATION = JPHGB,G37,075021;%%

\bibitem{topmass}
    [Tevatron Electroweak Working Group and CDF Collaboration and D0 Collab],
  %``Combination of CDF and D0 Results on the Mass of the Top Quark,''
  arXiv:1007.3178 [hep-ex].
  %%CITATION = ARXIV:1007.3178;%%

\bibitem{botmass}
  J.~H.~Kuhn, M.~Steinhauser and C.~Sturm,
  %``Heavy quark masses from sum rules in four-loop approximation,''
  Nucl.\ Phys.\  B {\bf 778} (2007) 192
  [arXiv:hep-ph/0702103];
  %%CITATION = NUPHA,B778,192;%%
%
  K.~G.~Chetyrkin {\it et al.}, 
%J.~H.~Kuhn, A.~Maier, P.~Maierhofer, P.~Marquard,M.~Steinhauser and C.~Sturm,
  %``Charm and Bottom Quark Masses: an Update,''
  Phys.\ Rev.\  D {\bf 80} (2009) 074010
  [arXiv:0907.2110 [hep-ph]];
  %%CITATION = PHRVA,D80,074010;%%
%
  J.~H.~Kuhn,
  %``Precise Charm- and Bottom-Quark Masses: Recent Updates,''
  PoS {\bf RADCOR2009} (2010) 035
  [arXiv:1001.5173 [hep-ph]].
  %%CITATION = POSCI,RADCOR2009,035;%%

\bibitem{atlas}
The ATLAS Collaboration, Conference Note ATLAS-CONF-2011-086.


\bibitem{bt}
  D.~Binosi and L.~Theussl,
  %``JaxoDraw: A graphical user interface for drawing Feynman diagrams,''
  Comput.\ Phys.\ Commun.\  {\bf 161} (2004) 76
  [arXiv:hep-ph/0309015].
  %%CITATION = CPHCB,161,76;%

\bibitem{FDS}
  B.~Field, S.~Dawson and J.~Smith,
  %``Scalar and pseudoscalar Higgs boson plus one jet production at the CERN LHC
  %and Tevatron,''
  Phys.\ Rev.\  D {\bf 69} (2004) 074013
  [arXiv:hep-ph/0311199].
  %%CITATION = PHRVA,D69,074013;%%

\end{thebibliography}
\end{document}